\documentclass[]{aa}
\usepackage{txfonts}
\usepackage {graphics}
\usepackage{placeins}
\usepackage {natbib}
\usepackage[figuresright]{rotating}
\bibpunct{(}{)}{;}{a}{}{,}
\usepackage{longtable,array,tabularx}

\begin{document}
\title{\object{Mrk 609}: resolving the circumnuclear structure with near-infrared integral field spectroscopy
\thanks{Based on ESO-VLT SINFONI science verification 60.A-9041(A) and Nobeyama Radio Observatory 45~m telescope observations.}}
\author{J. Zuther\inst{1}, C. Iserlohe\inst{1}, J.-U. Pott\inst{1,2}, T. Bertram\inst{1}, S. Fischer\inst{1}, W. Voges\inst{3}, G. Hasinger\inst{3}, A. Eckart\inst{1}}
\institute{I. Physikalisches Institut, Universit\"at zu K\"oln,
Z\"ulpicher Str. 77, 50937 K\"oln, Germany \and European Southern Observatory (ESO), Garching bei M\"unchen, Germany \and Max-Planck Institut f\"ur extraterrestrische Physik, Garching bei M\"unchen, Germany}
\offprints{J. Zuther, \email{zuther@ph1.uni-koeln.de}}
\date{Received  / Accepted }
\abstract
{}
{We present first results of near-infrared (NIR) $J$ and $H+K$ ESO-SINFONI integral field spectroscopy of the composite starburst/Seyfert 1.8 galaxy Mrk~609. The data were taken during the science verification period of SINFONI. We aim to investigate the morphology and excitation conditions within the central 2~kpc. Additional Nobeyama 45~m CO(1-0) data are presented, which we used to estimate the molecular gas mass. The source was selected from a sample of SDSS/ROSAT-based, X-ray bright AGN with redshifts of $0.03 < z < 1$ that are suitable for adaptive optics observations. This sample allows for a detailed study of the NIR properties of the nuclear and host environments with high spectral and spatial resolution.}
{Integral field spectroscopy with SINFONI delivers simultaneous spatial and spectral coverage of the circumnuclear environment. The NIR light is influenced less by dust extinction than by optical light and is sensitive to mass- dominating stellar populations. Furthermore, several NIR emission lines allow us to  distinguish between Seyfert and starburst activities.}
{Our NIR data reveal a complex emission-line morphology that is possibly associated with a nuclear bar seen in the reconstructed continuum images. The detections of [\ion{Si}{VI}] and a broad Pa$\alpha$ component are clear indicators of the presence of an accreting super-massive black hole at the center of Mrk~609. In agreement with previous observations, we find that the circumnuclear emission is not significantly extincted. Analysis of the high angular-resolution, molecular hydrogen emission and [\ion{Fe}{II}] emission reveals the LINER character of the nucleus. The large H$_2$ gas mass deduced from the CO(1-0) observation provides the fuel needed to feed the starburst and Seyfert activity in Mrk~609.}
{High angular resolution imaging spectroscopy provides an ideal tool for resolving the nuclear and starburst contributions in active galaxies. We show that Mrk~609 exhibits LINER features that appear to be hidden in visible/NIR spectra with larger apertures.}
\keywords{Galaxies: Seyfert -- Galaxies: starburst -- Galaxies: fundamental parameters -- Infrared: galaxies -- Galaxies: individual: Mrk 609}

\authorrunning{J. Zuther et al.}
\titlerunning{SINFONI's view of Mrk~609}
\maketitle

\section{Introduction}
\label{intro}

Stellar kinematical studies of nearby galaxies give grounds speculating that every galaxy harbors a super-massive black hole (SMBH) in its center \citep[e.g.][]{1995ARA&A..33..581K,2000ApJ...539L..13G,2002ApJ...574..740T}.
Numerous studies of active galactic nuclei (AGN) and quiescent galaxies have  furthermore revealed a correlation between the SMBHs and the bulges hosting them \citep[e.g.][]{1998AJ....115.2285M,2001Sci...294.2516P,2002ApJ...574..740T}.  These correlations indicate the existence of feedback mechanisms that regulate the coeval growth of SMBHs and their associated bulges.

In this paper we present results of an NIR study of the composite galaxy Mrk~609 \citep[Fig. \ref{fig:hst} and ][]{2006NewAR..49..508Z} with the new adaptive-optics (AO) assisted integral field spectrometer (IFS) SINFONI at the ESO VLT. For the first time, AO-assisted imaging spectroscopy on 8-10~m class telescopes allows for the simultaneous study of the morphology, chemical composition, and kinematics of the circumnuclear regions of AGN at an unprecedented spatial resolution.

\subsection{Fueling of nuclear activity}
\label{sec:fueling}
The presence of the Seyfert phenomenon is supposed to originate in the accretion of matter onto an SMBH in the centers of the galaxies. The fuel necessary for driving nuclear activity, which is composed of nuclear starbursts and  Seyfert-like activity, has to be transported from galactic scales ($\sim$10~kpc) down to nuclear scales of $\sim$10~pc. We are still far from understanding the detailed processes that lead to the dissipation of angular momentum needed for the gas and stars to fall towards the nuclear region. However, considerable theoretical, as well as observational, effort has been made toward understanding 
these processes \citep[e.g.][]{1990Natur.345..679S,1996IAUS..169..133C,2005Ap&SS.295...85K}. External and internal triggers of the fueling process can be distinguished. 

{\it External} triggers are related to the environment of galaxies and gravitational interaction. Non-axisymmetries, which can lead to loss of angular momentum, can  result from galaxy interactions. For ULIRGs, which show the most extreme case of infrared nuclear activity, there is intriguing evidence of a connection between galaxy interaction and nuclear activity \citep[cf. review by][~and references therein]{2005Ap&SS.295...85K}. Nevertheless, statistical studies show a significant fraction of nuclear active galaxies that are apparently free of any external trigger. Among the {\it internal} triggers, for example, \cite{1989Natur.338...45S} proposed a two-step process that is able to sweep the interstellar medium (ISM) via a stellar bar from large scales into a disk of several hundred pc in radius. In the second step, further instabilities (bar-within-bar) drive the material close to the nucleus until viscous processes take over the angular momentum transport. 

While extensive star formation in quiescent galaxies appears to be related to large-scale bars \citep[][~and references therein]{2005Ap&SS.295...85K}, the observational evidence of bar-related nuclear activity is not as clear for Seyfert galaxies. \cite{2002ApJ...567...97L} find a slight but significant increase in the galactic-bar fraction of active galaxies when compared to in-active galaxies. This does not appear to be the case for nuclear bars. On the other hand, there are a considerable number of AGN that show no signs of any bar or inactive galaxies that do possess bars. The overall lack of observational evidence of direct causal relationships between the presence of morphological asymmetries and the starburst/Seyfert activity might result from disregarding the correct spatial or time scales \citep{2005Ap&SS.295...85K}. To some extent the non-axisymmetries could occur on spatial scales that are currently not resolvable, or could be masked by dust or star formation \citep{2002ApJ...567...97L}. Considering lifetimes, the typical age for nearby spiral galaxies is $10^{10}$~yr. Only $\sim 10$\% of nearby spirals display Seyfert activity \citep[e.g.][]{1997ApJ...487..568H}. \cite{1988ApJ...325...74S} propose an evolutionary scenario in which the Seyfert phenomenon is a transient phase. If all galaxies undergo this phase, the small fraction of active galaxies translates into an AGN lifetime of about $10^9$~yr. But the non-stellar activity can also occur episodically on much shorter time scales \citep{2004cbhg.symp..169M}. The typical bar lifetime is $10^9$~yr \citep{2003AJ....126.1690C}, and the bar can trigger significant star formation within $10^8$~yr \citep{1999ApJ...516..660H}. Active circumnuclear star formation might, therefore, be related to the presence of a bar. But what happens with the bar when the Seyfert activity, under the assumption of bar- induced nuclear fueling, is episodic with a lifetime of about $10^4$~yr? Further work has clearly to be done, primarily aimed at the kinematics and dynamics of the very central region of active and inactive galaxies.

The recent finding of a stellar surface mass density criterion for the separation of bulge-dominated and disk-dominated galaxies by \cite{2003MNRAS.346.1055K} can shed some light on the problem. The authors find that bulges dominate in stellar masses $\ga 3\times 10^{10} M_\odot$ corresponding to a mean stellar surface density of $\mu_\star\ga 3\times 10^{8} M_\odot$~kpc$^{-2}$. \cite{2004ApJ...613..109H} find that this stellar surface density is also the critical surface density for the onset of significant Seyfert activity. These results can be compared with the critical stellar surface density necessary for the onset of bar instability \citep{2004ApJ...612L..17W}. It turns out that this surface density ($\mu > 1.7\times 10^8 M_\odot$~kpc$^{-2}$) is very similar to the one found by \cite{2004ApJ...613..109H} for nuclear activity. One expects that the stellar surface density increases once a bar is formed, since the subsequent gas infall towards the central regions will be accompanied by star formation. This is seen in both observations and in simulations and, thus, can explain the factor of about 2 below the observed stellar surface density for bulge dominance.

\subsection{Starburst/Seyfert composite galaxies}
\label{sec:composites}
A subgroup of AGN, the \emph{starburst/Seyfert composite} galaxies \citep{1996ApJS..106..341M}, appears to be best suited to studying the starburst-AGN connection, since the AGN and starburst components present themselves at the same level of activity \citep{2005ApJ...631..707P}. These objects can be characterized by optical spectra that are dominated by starburst features, while the X-ray luminosity and its variability are typical of Seyfert galaxies.
The former property is based on the emission-line diagnostic diagrams by \citet{1987ApJS...63..295V}. Close inspection of the optical spectra often reveals some weak Seyfert-like features, e.g. [\ion{O}{III}] being significantly broader than all other narrow lines or a weak broad H$\alpha$ component. There is a resemblance to narrow-line X-ray galaxies \citep[NLXG, e.g.][]{1995MNRAS.276..315B}, which also show spectra of a composite nature. Their soft X-ray spectra are hard, i.e. flat \citep{1996MNRAS.282..295A}, but it is still not clear how this strong and hard X-ray emission can be reconciled with the weak/absent optical Seyfert characteristics. The faintness of these objects in the X-ray, as well as in the optical domain, has not allowed us to study them in detail so far.

Near-infrared (NIR) imaging spectroscopy has considerable advantages over visible wavelengths: Besides the much smaller dust extinction, there are a number of NIR diagnostic lines (in emission and in absorption) to probe the excitation mechanisms and stellar content in galaxies. Among these are hydrogen recombination lines, rotational/vibrational transitions of H$_2$, stellar features like the CO(2-0) and CO(6-3) absorption band heads, and forbidden lines like [\ion{Fe}{II}] and [\ion{Si}{VI}] \citep{1999AJ....117..111H,1994ApJ...427..777M,1994A&A...291...18M}. 

\subsection{Mrk~609}
\label{sec:mrk609}

Mrk~609 shows a mixture of nuclear star formation and Seyfert activity and appears to be a showcase for studying the nuclear activity with IFSs in order to probe the coeval existence of these phenomena.

\begin{figure}[h!]
\begin{center}
\resizebox{\hsize}{!}{\includegraphics{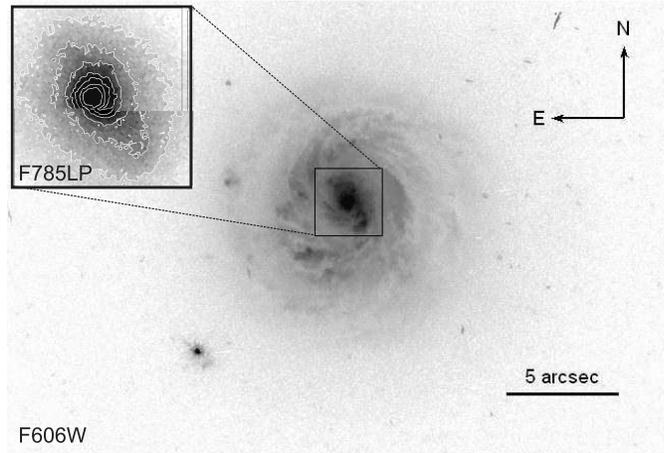}}
\end{center}
\caption{HST F606W ($\sim$0.6~$\mu$m) image of Mrk~609 \citep{1998ApJS..117...25M}. The black box indicates the SINFONI field of view of $3\times3$~arcsec$^2$. The upper left inset is an HST F785LP ($\sim$0.9~$\mu$m) image of the nuclear $3\times 3$~arcsec$^2$ \citep{1996ApJ...466..713N}.  Compare with Figs. \ref{fig:hk_extended}~(d), \ref{fig:j_band}~(a), and the structure image (Fig. 2) of \cite{2006AJ....132..321D}.}
\label{fig:hst}
\end{figure}

There have been several observations of Mrk~609 (Figs. \ref{fig:hst} and \ref{fig:sed} for the SED). It is a member of the original sample of 
\citet{1981ApJ...249..462O}, with which the role of reddening in Seyfert 1.8/1.9 galaxies was studied. These observations were followed up by X-ray, ultraviolet, optical, and infrared studies \citep{2002MNRAS.336..714P, 1988ApJ...332..172R,1990ApJ...355...88G,1985ApJ...298..614R}, refining the classification of Mrk~609 as a starburst/Seyfert composite. The UV/optical line ratios and the X-ray spectrum furthermore indicate a small extinction towards the nucleus. \citet{1988ApJ...332..172R} and \citet{1990ApJ...355...88G} suggest that smaller-than-normal optical depths and lower ionization are responsible for the large Balmer decrement found in the optical spectrum (Fig. \ref{fig:sdss_spec}).

Optical \citep[$\sim$0.6~$\mu$m;][]{1996ApJ...466..713N} and NIR \citep[$\sim$0.9~$\mu$m][]{2003AJ....126.1690C} imaging observations of Mrk~609 with the Hubble Space Telescope (HST) have been used to address questions regarding the extension of the scattering medium in Seyfert 2 galaxies and the dependence of AGN fueling on the host morphology (Fig. \ref{fig:hst}). \citeauthor{1996ApJ...466..713N} studied a sample of Seyfert and non-Seyfert Markarian galaxies and  found that Seyfert 1.8 and 1.9 show extended nuclear light profiles similar to Seyfert 2 galaxies. The energetics of the  (circum-) nuclear region of Seyfert 2s is strongly influenced by starburst activity. In combination with some scattering screen, this provides the extinction responsible for the observed shape of the light profiles. Seyfert 1.5 galaxies, on the other hand, have nuclear profiles typical of Seyfert 1 galaxies, which are unresolved and bright. Mrk~609 shows a strong Seyfert 1-like nucleus, but it appears to be slightly extended. Furthermore, weak broad H-recombination lines have been detected in optical spectra, which result in an intermediate Seyfert 1.5-1.8 classification \citep{1990ApJ...355...88G,1981ApJ...249..462O}. This work supports the above findings. 

\citeauthor{2003AJ....126.1690C} and \cite{2006AJ....132..321D} find a larger bar fraction for narrow-line Seyfert 1 (NLS1) galaxies than for broad-line Seyferts (BLS1). NLS1 are believed to be systems accreting close-to or above the Eddington limit. The active state requires an efficient fuel supply, which can be supported by stellar bars. As a BLS1, the large-scale optical morphology of Mrk~609 appears flocculent and shows no signs of a bar \citep{2003AJ....126.1690C, 2006AJ....132..321D} (Fig. \ref{fig:hst}). The nuclear morphology, however, deviates from a circular symmetric appearance. The NIR image of \citeauthor{1996ApJ...466..713N} (see inset in Fig. \ref{fig:hst}), which is less influenced by dust extinction than the optical image, reflects this situation. Two spiral arms turn out to connect to an elongated nuclear structure (northeast-southwest direction), probably resembling a nuclear stellar bar. In a recent numerical study, \cite{2000ApJ...528..677E} find that a grand-design nuclear two-arm spiral structure can be formed in the presence of a weak stellar bar. Only depending on the nuclear mass and the gas sound speed, such structures can allow for the inward mass transport beyond the inner Lindblad resonance towards radii of 50-100~pc. In respect thereof \cite{2006AJ....132..321D} find that especially the class of flocculent spirals avoids the presence of large-scale bars, as appears to be the case for Mrk~609. 

The paper is structured in the following manner. In Sect. \ref{sec:obs} we describe the observations and data reduction. After a detailed analysis of the NIR and mm spectra, we discuss the results within this context of nuclear fueling in Sect. \ref{sec:results}. In Sect. \ref{sec:summary} we summarize the presented work and give an outlook. Unless otherwise noted, we use a cosmology with $H_0=70$~km~s$^{-1}$~Mpc$^{-1}$, $\Omega_m=0.3$, and $\Omega_\Lambda=0.7$ \citep{2003ApJS..148..175S} throughout this paper. 
\section{Observations and data reduction}
\label{sec:obs}
\subsection{Near-infrared data}
\label{sec:nir_reduction}
The observations were carried out during the science verification phase\footnote{http://www.eso.org/science/vltsv/sinfonisv/xrayagn.html} (October 2004) of SINFONI, the new AO-assisted integral field spectrometer mounted at Yepun, Unit Telescope 4 of the ESO Very Large Telescope in Chile \citep{2003SPIE.4841.1548E}. The AO guiding was carried out on the nucleus of Mrk~609. The average seeing was around 0\farcs7. The 100~mas pixel scale with a field-of-view (FOV) of 3\arcsec$\times$3\arcsec was used. The 2D image on the sky was sliced by small mirrors into 32 slitlets, which then were reimaged onto a pseudoslit and dispersed onto a 2k$\times$2k detector.
The observations covered $J$ and $H+K$ bands with integration times of 5 minutes and 30 minutes, respectively. Dispersion was achieved with the $J$ and $H+K$ gratings at a spectral resolution of $R_J\sim 2000$ and $R_{H+K}\sim 1500$. Successive target (T) and sky (S) observations (...TSST...) were acquired to produce sky-subtracted frames.

The reduction and reconstruction of the 3D cubes were carried out using the MPE reduction software \emph{spred 3.6} \footnote{Kindly provided by Matthew Horrobin (MPE, Garching).}, IDL, and QFitsView\footnote{Written by Thomas Ott (MPE, Garching); http://www.mpe.mpg.de/$\sim$ott/QFitsView}. Bad pixel, cosmic rays, and flat field corrections were applied to the 2D raw frames. The 3D cubes were reconstructed using calibration frames for the slitlet distances and the light dispersion.

Intermediate standard-star observations (near in both time and airmass to the target exposures) of the G2V star HIP-021070 were used to correct for strong atmospheric (telluric) absorptions. They can be minimized by dividing the science target spectrum by the standard star spectrum. This step introduces features into the resulting spectrum that are intrinsic to the standard star and that can be accounted for by multiplication by the atmospheric-transmission-corrected solar spectrum. Since the standard star was saturated on the detector (see below in the context of flux calibration), we extracted the spectrum with telluric absorptions from the wings of the 2D image. The G2V spectral characteristics were removed by multiplying by the well known, high signal-to-noise solar spectrum \citep[as provided by][]{1996AJ....111..537M}. In the range of reduced atmospheric transmission (as indicated in Fig. \ref{fig:hk_spec}), which is not covered by the \citeauthor{1996AJ....111..537M} spectrum, we interpolated the solar spectrum with a blackbody, because the original solar atlas did not show any intrinsic features in that range \citep{1996ApJS..106..165W}.

Flux calibration is problematic, since the calibration-star observations saturated the detector. In order to approach a flux calibration, ignoring the fact of variability (up to 30\%) of Mrk~609 found by \citet{1985ApJ...298..614R}, we used the 2MASS $J$, $H$, and $Ks$ fluxes. We measured the fluxes within a 3\arcsec diameter aperture in the 2MASS atlas images and applied the fluxes to the $3\arcsec\times 3\arcsec$ SINFONI FOV. It has to be noted that this method has the drawback of ignoring the differences in spatial resolution. We expect an accuracy of the absolute flux to be around 40\%, but this is not of any significant importance, because we are primarily interested in measuring line ratios. For these, in principle, an absolute flux calibration is not necessary. We converted the magnitudes to flux densities using the Spitzer space telescope magnitude to flux density converter\footnote{ (http://ssc.spitzer.caltech.edu/tools/magtojy/)}. For $J$ we measured $J_\mathrm{2MASS}=13.68$, which corresponds to $f_{J\mathrm{2MASS}}=1.06\times 10^{-14}$~W~m$^{-2}$$\mu$m$^{-1}$ at $\lambda_\mathrm{eff}(J_\mathrm{2MASS})=1.235$. For the $H+K$ band we used the flux density at the effective wavelength of the $Ks$ band. We measured $Ks_\mathrm{2MASS}=12.50$, which corresponds to $f_{Ks\mathrm{2MASS}}=4.29\times 10^{-15}$~W~m$^{-2}$$\mu$m$^{-1}$ at $\lambda_\mathrm{eff}(Ks_\mathrm{2MASS})=2.159$. The total flux within the SINFONI FOV was scaled according to these values. Because of the low value of $E(B-V)=0.056$ \citep{1998ApJ...500..525S}, no correction for Galactic extinction is applied.

From the final reconstructed 3D cube we extracted spectra and emission line maps. The emission line maps presented in Figs. \ref{fig:hk_extended}, \ref{fig:hk_nuclear}, and \ref{fig:j_band} were created by summing up the flux of the emission line and then subtracting the average flux density of the continuum left and right of the line. The features seen in the Pa$\alpha$ map define the regions where we extracted spectra (Figs. \ref{fig:hk_spec}, \ref{fig:pa_a_nuc}, \ref{fig:j_spec}, and \ref{fig:pa_b_nuc}). The radius of the extraction regions is 5 pixels (0\farcs 25). 

\subsection{Millimeter data}
During an observing campaign of cluster galaxies with the Nobeyama 45m telescope in March 2005 we additionally observed Mrk~609 in $^{12}$CO(1-0). We used the autocorrelator as backend and an integration time of 45 minutes. The measured antenna temperatures were transformed to a main-beam temperature using a main-beam efficiency of 0.38. The baseline-subtracted CO(1-0) spectrum is presented in Fig. \ref{fig:nobeyama}.

\section{Results and discussion}
\label{sec:results}
In this section we describe the results of the spatially resolved spectroscopy of the circumnuclear environment of Mrk~609 and the CO(1-0) observation.

\subsection{Tracing the continuum and emission line gas}
\begin{figure*}[h!]
\begin{center}
\resizebox{\hsize}{!}{\includegraphics{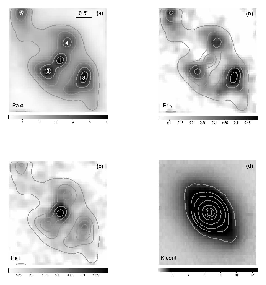}}
\caption{$H+K$ emission line maps of recombination lines. They were created with QFITSVIEW by summing the flux over the emission line and subtracting the median value of the neighboring continuum (a-c). All images were smoothed with a 3- pixel Gaussian. The values of the color bars are in units of $10^{-19}$~W~m$^{-2}$. (a) Narrow Pa$\alpha$; contours are calculated relative to the peak flux (90, 70, 50, 30\%). (b) Br$\gamma$ overlayed with Pa$\alpha$ contours. (c) \ion{He}{I}~2.059~$\mu$m overlayed with Pa$\alpha$ contours. (d) Continuum image in a region without emission lines around 2.2~$\mu$m. The panels are $3\arcsec\times3\arcsec$ each. North is up and east is left. Regions are indicated in panel (a) from which we extracted cummulative spectra. Regions 2 and 4 have a projected distance of about 420~pc, region 3 $\approx 630$~pc, and region 5 $\approx 1.3$~kpc to the center.}
\label{fig:hk_extended}
\end{center}
\end{figure*}

\begin{figure*}[h!]
\begin{center}
\resizebox{\hsize}{!}{\includegraphics{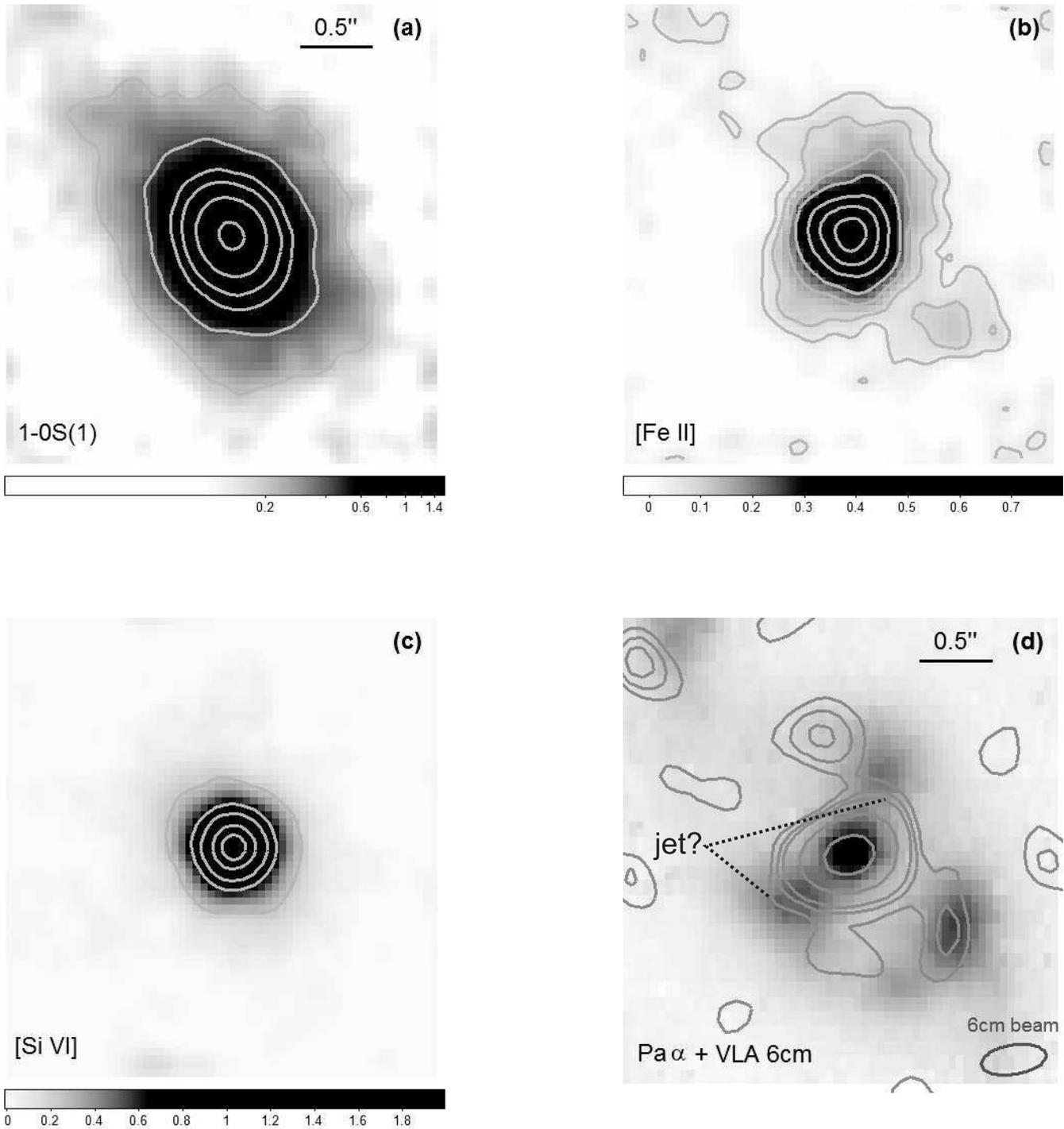}}
\caption{$H+K$ emission line maps (using the same procedure as for Fig. \ref{fig:hk_extended}) of: (a) 1-0S(1) (contours: 90, 70, 50, 30, and 10\% of peak flux), (b) [\ion{Fe}{II}]~1.664~$\mu$m (contours: 90, 70, 50, 30, 20, 10, and 5\% of peak flux), and (c) [\ion{Si}{VI}] (contours: 90, 70, 50, 30, and 10\% of peak flux). The colorbar values are in units of $10^{-19}$~W~m$^{-2}$. Panel (d) displays an overlay of the Pa$\alpha$ map from Fig. \ref{fig:hk_extended}(a) with the 6~cm VLA map in (magenta) contours  \citep[from][]{1984ApJ...278..544U}. The blue ellipse in the lower right corner indicates the 6~cm beam size. The panels are $3\arcsec\times3\arcsec$ each. North is up and east is left.}
\label{fig:hk_nuclear}
\end{center}
\end{figure*}

\begin{figure*}[h!]
\begin{center}
\resizebox{\hsize}{!}{\includegraphics{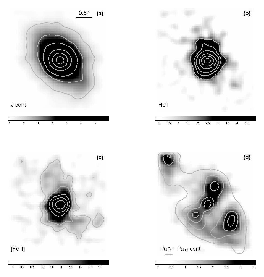}}
\caption{$J$-band continuum and emission line maps of: (a) 1.25~$\mu$m continuum (contours: 90, 70, 50, 30, 20, and 10\% of peak flux), (b) \ion{He}{I}~1.083~$\mu$m (contours: 90, 70, 50, 30, 20, and 10\% of peak flux), (c) \ion{Fe}{II}~1.257~$\mu$m (contours: 90, 70, 50, 30, and 15\% of peak flux), and (d) Pa$\beta$ overlayed with Pa$\alpha$ in contours (see Fig. \ref{fig:hk_extended}(a)). Color bar values are in units of $10^{-19}$~W~m$^{-2}$. The panels are $3\times3$ arcsec$^2$ each. North is up and east is left.}
\label{fig:j_band}
\end{center}
\end{figure*}

\begin{figure}
\begin{center}
\resizebox{\hsize}{!}{\includegraphics{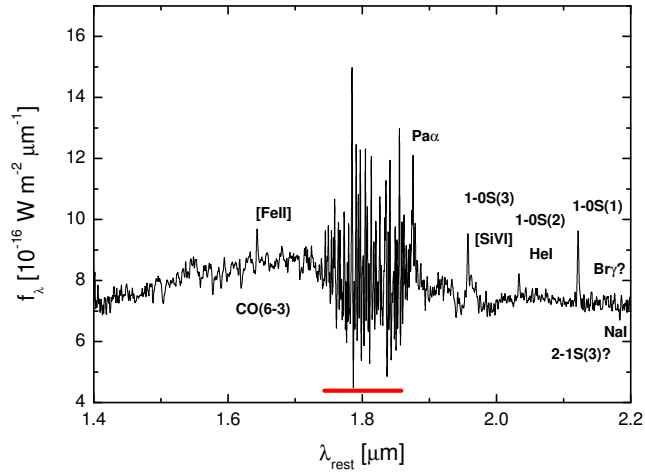}}
\caption{$H+K$ rest-frame spectrum extracted from a 5-pixel radius aperture centered on the nucleus (region 1 in Fig. \ref{fig:hk_extended} (a)). Prominent lines are indicated (cf. Table \ref{tab:linefluxes}). The high noise around 1.8~$\mu$m, indicated by the thick bar, is due to reduced atmospheric transmission between the $H$ and the $K$ bands.}
\label{fig:hk_spec}
\end{center}
\end{figure}

The SINFONI imaging spectroscopy allows us to study the inner 2~kpc of Mrk~609 at a spatial resolution of about 270~pc. We identified several emission and absorption features in the individual $J$ and $H+K$ spectra across the FOV (see  Figs. \ref{fig:hk_spec} and \ref{fig:j_spec}). Among these are hydrogen recombination lines, rotational/vibrational lines of molecular hydrogen, forbidden transitions of [\ion{Fe}{II}] and [\ion{Si}{VI}], as well as stellar CO(6-3) and CO(2-0) absorption. From selected emission lines we produced continuum-subtracted line maps (Figs. \ref{fig:hk_extended}, \ref{fig:hk_nuclear}, and \ref{fig:j_band}).

In order to assess the spatial resolution we used the nuclear Pa$\alpha$ emission. Fitting a 2D Gaussian to the nuclear Pa$\alpha$ gives a rough estimate for the spatial resolution, because no other point-source observations were available for our observations. We found an FWHM of about 8 pixel ($\sim$0\farcs 4) for the minor axis, while the major axis is clearly extended. We expect this to be a reasonable measure of the point-spread function at all wavelengths. This size corresponds to a linear scale of about 270~pc ($1\arcsec\approx 680$~pc) at the redshift of $z=0.0345$.

The morphologies of the hydrogen recombination line maps cleary differ from the continuum maps. The Pa$\alpha$ map reveals five emission peaks that are indicated in Fig. \ref{fig:hk_extended} (a). The continuum emission presented in panel (d) of Fig. \ref{fig:hk_extended} shows an elliptical structure extending in a northeast/southwest direction and connecting to the large-scale spiral arms  visible in the HST image (Fig. \ref{fig:hst}). This morphology is similar to that of the nuclear stellar bars found in several other galaxies \citep[e.g.][]{2002AJ....124...65E,2003ApJ...589..774M}. Results of N-body simulations carried out by \cite{2005MNRAS.358..305P}, in which stellar particles move on (quasi-) periodic orbits in a bar potential, also resemble the nuclear continuum morphology of Mrk~609. 

The Pa$\alpha$ peaks are apparently aligned with the barred continuum. One peak is centered on the nucleus. The four other peaks lie at the tip of the major axis where the bar connects to the spiral arms (regions 3 and 5) and on the minor axis (regions 2 and 4). In addition to the clumpy Pa$\alpha$ emission a fainter and smoother Pa$\alpha$ component is visible, which extends between the distinct regions. There is a striking similarity between this configuration and the observation of \cite{2003MNRAS.345.1297E}, who detected a nuclear two-arm gas spiral in the early type galaxy NGC~2974. This spiral structure ($\sim 200$~pc radius) is embedded in a $\sim 500$~pc radius nuclear stellar bar. These dimensions are roughly the same (within a factor of 2) as for Mrk~609 (Fig. \ref{fig:hk_extended} (a) and (d)). Simulations show that gaseous material can be transported along these spiral arms towards the central region \citep{2000ApJ...528..677E}. Mrk~609 has also been studied at 6~cm wavelengths \citep{1984ApJ...278..544U}. Figure \ref{fig:hk_nuclear} (d) displays the Pa$\alpha$ map overlayed with the radio contours. The radio emission appears to be slightly resolved and it also coincides with the recombination line emission. Accordingly, the 6~cm radio continuum traces the zones of ongoing star formation. As indicated in Fig. \ref{fig:hk_extended} (d), there is an extension towards regions 2 and 4, which could also be interpreted as a jet component. Regions 2 and 4, therefore, might be excited by the radio jet impinging on the inter-stellar medium. In this case the nuclear spiral scenario has to be altered. See Sects. \ref{sec:kinematics} and \ref{sec:connection} for further discussion.

\begin{figure}
\begin{center}
\resizebox{\hsize}{!}{\includegraphics[angle=90]{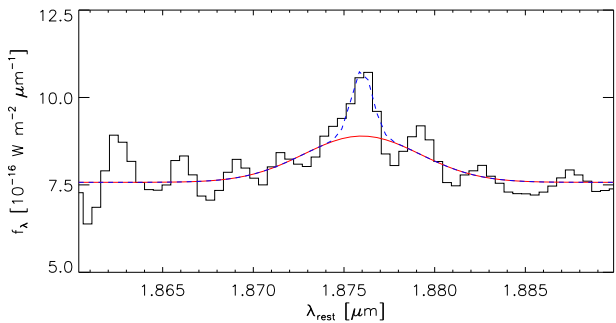}}
\caption{Rest-frame spectrum around Pa$\alpha$, extracted from the nuclear 5 pixel radius aperture. The spectrum has been smoothed with a 3-pixel boxcar. The blue dashed and red solid lines represent Gaussian fits to the narrow and broad components of Pa$\alpha$.}
\label{fig:pa_a_nuc}
\end{center}
\end{figure}

\begin{figure}
\begin{center}
\resizebox{\hsize}{!}{\includegraphics{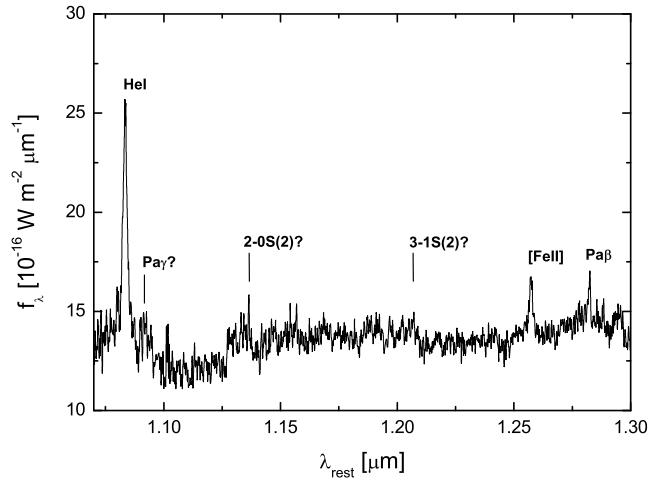}}
\caption{$J$ rest-frame spectrum extracted from a 5 pixel radius aperture centered on the nucleus (region 1 in Fig. \ref{fig:hk_extended} (a)). Prominent lines are indicated (cf. Table \ref{tab:linefluxes}).}
\label{fig:j_spec}
\end{center}
\end{figure}

\begin{figure}
\begin{center}
\resizebox{\hsize}{!}{\includegraphics[angle=90]{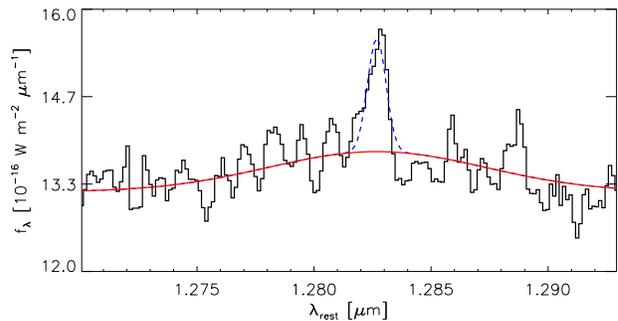}}
\caption{Rest frame spectrum around Pa$\beta$, extracted from the central 5 pixel radius aperture. The spectrum has been smoothed with a 3-pixel boxcar and the result of the fit of a narrow (blue dashed curve) and a broad (red solid curve) Gaussian component with a common line center is shown.}
\label{fig:pa_b_nuc}
\end{center}
\end{figure}

Next, we extracted 1D spectra from the $H+K$ and $J$ bands for each of the five regions by summing up the individual spectra within apertures of 5 pixel radius, centered on the Pa$\alpha$ emission peaks (see Figs. \ref{fig:hk_spec} and \ref{fig:j_spec}). The apertures sizes have been chosen to represent the angular resolution of the observation. In the resulting spectra we measured line fluxes for individual lines by direct measurement. Line fluxes and FWHM of the line profiles are given in Tab. \ref{tab:linefluxes}. The upper part of the table corresponds to the $J$ part of the spectrum, the lower one to the $H+K$ part. Fluxes are given in units of $10^{-19}$~W~m$^{-2}$ and observed line widths are given in km~s$^{-1}$. Errors of the line fluxes were estimated from the noise of the neighboring continuum alone. One has to keep in mind that errors introduced by the flux calibration become relevant when comparing line fluxes between bands ($J$ and $H+K$) or when using absolute fluxes (see discussion about calibration in Sect. \ref{sec:nir_reduction}). The measurement of the FWHM depends on the shape of the continuum. We estimate its error to be on the order of 30\%. In the case of Pa$\alpha$ and Pa$\beta$ we fitted a broad and narrow Gaussian component with a common line center (Figs. \ref{fig:pa_a_nuc} and \ref{fig:pa_b_nuc}). The measurement of the broad components is quite intricate, since these lines are located in problematic regions of the spectrum. For example, Pa$\alpha$ is located at the red end of the atmospheric transmission minimum ( $\sim1.8\mu$m), and the spectrum is very noisy around this feature. As for Pa$\beta$ and for \ion{He}{I}, they are located at the beginning and at the end of the $J$-band spectrum and are also influenced by increased noise. For the nucelar 1-0S(3)/[\ion{Si}{VI}] complex, we fitted two Gaussians and a telluric absorption component. A fit without the telluric component recovers the 1-0S(3) flux and linewidth within the errors given in Table \ref{tab:linefluxes}. In the case of the [\ion{Si}{VI}], the flux and linewidth are considerably reduced (by about 60\% and 30\%, respectively). Thus, primarily the properties of [\ion{Si}{VI}] and not of 1-0S(3) are influenced. Since we only state the presence of the former as an indicator of nuclear activity and do not consider it any further, we neglect the telluric absorption for the 1-0S(3) measurement.

In contrast to H$\alpha$ and H$\beta$ at visible wavelengths  (Fig.\ref{fig:sdss_spec}), Pa$\alpha$ (Fig. \ref{fig:pa_a_nuc}) and Pa$\beta$ (Fig. \ref{fig:pa_b_nuc}) clearly show a broad component with an FWHM of about 3000~kms$^{-1}$ (after correcting for the instrumental spectral resolution). This supports the Seyfert classification of Mrk~609 as deduced from the X-ray properties \citep{2002MNRAS.336..714P}. In the $J$-band spectrum, there is also evidence of a broad \ion{He}{I} 1.083~$\mu$m component, but the low signal-to-noise ratio prevents us from decomposing broad and narrow components. Therefore, we only state the flux from direct measurement of the narrow component in Table \ref{tab:linefluxes}. The broad component also agrees with the line width of the broad Ly$\alpha$ ($\sim 3600$~km~s$^{-1}$, but blended with \ion{N}{V}) found by \cite{1988ApJ...332..172R}, when taking the International Ultraviolet Explorer (IUE) spectral resolution of 7~\AA~ into account. The coronal [\ion{Si}{VI}] emission visible in the nuclear spectrum (Figs. \ref{fig:hk_spec} and \ref{fig:hk_nuclear}(c) for the line map) is another common feature of the extreme energetics in AGN \citep{1994A&A...291...18M,2005MNRAS.364L..28P}.

Using the unreddened case-B line ratio of Pa$\alpha$/H$\beta=0.332$ \citep{1989agna.book.....O}, we find that the broad component should be detectable in the visible; but as mentioned in the introduction, Mrk~609 exhibits unusual line ratios. Intrinsic extinction can be probed by comparing the ratios of hydrogen recombination lines with case-B values from \cite{1989agna.book.....O}. For $T=10^{4}$~K, $n_e=10^4$~cm$^{-3}$, and no extinction, we expect  $\mathrm{Pa}\alpha/\mathrm{Br}\gamma=12.4$. Table \ref{tab:ratios} lists this ratio among others for the regions discussed here. Within the measurement uncertainties, the Pa$\alpha$/Br$\gamma$ ratios are consistent with no intrinsic extinction. There is in fact a trend of the ratios being a bit larger then the case-B value. This is especially the case for the nuclear region, where the corresponding ratios of the broad components also show the same behavior. This result agrees with low H column densities derived from the X-ray spectra \citep{2002MNRAS.336..714P} and the unusual line ratios found in the UV/optical by \cite{1988ApJ...332..172R}, who explained these ratios with low optical depths and low ionization parameters. 

The 1.66~$\mu$m/2.16~$\mu$m~continuum color map (Fig. \ref{fig:reddening}) is sensitive to reddening caused by warm dust. The map shows reddening along the major bar axis, with peaks at the nucleus and at the tip of the bar, meeting the spiral arms. Previous studies of the NIR continuum of Seyfert galaxies concluded that their $J-H$ and $H-K$ nuclear colors can be explained by a mixture of a stellar component and a warm/hot dust component in emission \citep[e.g.][]{1982MNRAS.199..943H,1985MNRAS.214..429G,1992MNRAS.256..149K,1996MNRAS.278..902A}.
Using the $H$ and $K$ zero points from Table 3.4 in \cite{1999hia..book.....G}, the $H-K$ colors range from 0.6 in the outer regions up to 1.0 at the nucleus. 
These are compatible with warm ($\sim 500$~K) dust emission and small extinction \cite[cf.][]{1985MNRAS.214..429G}.
As \cite{1996ApJ...458..132C} and \cite{2003ApJ...599L..21F} furthermore show, the effects of dust in starbursts can often be fairly well modelled by a homogeneous or clumpy/turbulent foreground screen. Since the extinction curve is very flat in the NIR \citep{1989isa..book...93D} and since the wavelengths of the lines used in the present analysis have small separations, the differential reddening between these lines will be small. Therefore, the $H-K$ reddening will only have a neglegible influence on the measured line ratios. 

Due to the uncertainties in the absolute calibration, where we do not have reliable knowledge of the host galaxy and nuclear contribution in the measuring aperture, we will not use the Pa$\beta$/Pa$\beta$ ratio as an indicator for extinction. Also the $J$ band spectrum shows no significant extinction. If we assume the same case-B conditions as above, we would expect not to detect the Pa$\gamma$ line (close to \ion{He}{I}~1.0883~$\mu$m). Figure \ref{fig:j_band}, however, shows faint signs of Pa$\gamma$, which points towards the same conclusion of small extinction around the nucleus.

Pa$\beta$, Br$\gamma$, and \ion{He}{I}~2.058~$\mu$m exhibit the same spatial distribution (Fig \ref{fig:hk_extended} (b), (c), and Fig. \ref{fig:j_band} (d)), and their intrinsic line widths are comparable and are spectroscopically unresolved. This indicates that the lines arise in the same parcels of gas. Note that on the nucleus Br$\gamma$ is only measured as an upper limit and \ion{He}{I}~2.058~$\mu$m only has a narrow component. On the other hand \ion{He}{I}~1.0883~$\mu$m is just concentrated on the nucleus (Fig. \ref{fig:j_band} (b)) and has both a broad component and a narrow one. 

\begin{table*}
\caption{Emission line fluxes for the 5 regions identified in the Pa$\alpha$ 
line map (Fig. \ref{fig:hk_extended} (a)).} 
\label{tab:linefluxes}
\centering
\begin{tabular}{ccccccccccc} 
\hline\hline             
Line & \multicolumn{2}{c}{Region 1} & \multicolumn{2}{c}{Region 2} & \multicolumn{2}{c}{Region 3} & \multicolumn{2}{c}{Region 4} & \multicolumn{2}{c}{Region 5}\\
$\lambda$ [$\mu$m]& Flux & FWHM & Flux & FWHM & Flux & FWHM & Flux & FWHM & Flux & FWHM\\
\hline
\ion{He}{I}~$\lambda 1.0833$ narrow & $31.8\pm 10.6$      & 540  &  ---          & --- & ---          & --- & ---           & ---   & ---               & --- \\
$[$\ion{Fe}{II}$]$~$\lambda 1.257$                 & $5.1\pm 1.0$       & 439  & $1.3\pm 0.3$  & 327& $0.7\pm 0.1$  & 160 & $1.3\pm 0.2$  & 303   & ---               & --- \\
Pa$\beta$ narrow$^{\mathrm{a}}$                    & $1.9\pm 0.6$       & 214  & $1.7\pm 0.1$  & 150& $2.2\pm 0.1$  & 163 & $2.2\pm 0.2$  & 183   & ---               & --- \\
Pa$\beta$ broad$^{\mathrm{a}}$                     & $15.0\pm 7.0$      & 3400 &  ---          & ---                & --- & ---           & ---   & --- & ---         & --- \\
\hline
$[$\ion{Fe}{II}$]$~$\lambda 1.644$$^{\mathrm{c}}$  & $2.7\pm 0.2$       & 410   & $0.6\pm 0.1$ & 240 & $0.3\pm 0.1$  & 200 & $0.5\pm 0.1 $ & 260   & $0.07\pm 0.01$   & 200 \\
Pa$\alpha$ narrow                                  & $5.3\pm 0.6$       & 300   & $7.9\pm 0.2$ & 270 & $7.0\pm 0.1$  & 200 & $5.4\pm 0.2$  & 200   & $0.22\pm 0.01$   & 210 \\ 
Pa$\alpha$ broad                                   & $16.4\pm 5.6$      & 3000  & ---          & --- & ---           & --- & ---           &  ---  & ---              & --- \\
1-0S(3)$^{\mathrm{b,c}}$                           & $5.0\pm 0.8$       & 317   & $1.1\pm 0.2$ & 371 & $0.5\pm 0.1$  & 345 & $0.6\pm 0.2$  & 471   & $0.05\pm 0.02$   & 230 \\
$[\ion{Si}{VI}]$$^{\mathrm{b,c}}$                  & $16.4\pm 4.2$      & 2130  &  ---         & --- & ---           & --- & ---           &  ---  & ---              & --- \\
1-0S(2)                                            & $2.1\pm 0.4$       & 308   & $0.6\pm 0.1$ & 366 & $0.20\pm 0.06$& 321 & $0.4\pm 0.1$  & 333   & $0.06\pm 0.02$   & 450 \\
\ion{He}{I}~$\lambda 2.058$                        & 0.3$^{\mathrm{d}}$ & 300 & $0.3\pm 0.1$ & 338 & $0.20\pm 0.05$& 200 & $0.20\pm 0.05$& 200     & $0.07\pm 0.01$   & 200 \\
2-1S(3)$^{\mathrm{e}}$                             & 0.2                & 317 & 0.1          & 322 & 0.1           & 312 & 0.1           & 316     & 0.02             & 300 \\ 
1-0S(1)                                            & $6.4\pm 0.4$       & 330 & $1.6\pm 0.1$ & 375 & $0.6\pm 0.04$ & 270 & $0.9\pm 0.1 $ & 300     & $0.10\pm 0.01$   & 300 \\ 
Br$\gamma$                                         & 0.37$^{\mathrm{d}}$& 300 & $0.5\pm 0.1$ & 267 & $0.6\pm 0.1$  & 200 & $0.4\pm 0.1 $ & 200     & $0.18\pm 0.01$   & 200 \\
\hline
\end{tabular}
\begin{list}{}{}
\item[$^{\mathrm{a}}$] Decomposition using two Gaussian components.
\item[$^{\mathrm{b}}$] 1-0S(3) and $[$\ion{Si}{VI}$]$ are decomposed using two Gaussian components.
\item[$^{\mathrm{c}}$] Result strongly influenced by telluric absorption.
\item[$^{\mathrm{d}}$] Upper limit, assuming the FWHM of the narrow Pa$\alpha$ component.
\item[$^{\mathrm{e}}$] Upper limit, assuming the average FWHM of the other molecular hydrogen lines.
\end{list}
\vfill
\end{table*}

\begin{table}[ht!]
\caption{Ratios of prominent emission lines using the narrow components.}
\begin{center}
\begin{tabular}{ c c c c}
\hline
\hline
Region & $\frac{\mathrm{Pa}\alpha}{\mathrm{Br}\gamma}$ & $\frac{\mathrm{H}_2 2.121\mu\mathrm{m}}{\mathrm{Br}\gamma}$ & $\frac{[\ion{Fe}{II}]1.257\mu\mathrm{m}}{\mathrm{Pa}\beta}$\\
\hline
1 &$>14.3$ ($>13.7$)$^{\mathrm{a}}$& $>17.3$ & $3\pm 1$ \\
2 & $15\pm 3$&$3.2\pm 0.7$& $0.8\pm 0.2$\\
3 & $12\pm 2$& $1.0\pm 0.2$& $0.32\pm 0.04$\\
4 & $14\pm 3$& $2.3\pm 0.6$& $0.5\pm 0.1$\\
5 & $12.2\pm 0.1$& $0.6\pm 0.2$ &--\\
\hline
\end{tabular}
\end{center}
\label{tab:ratios}
\begin{list}{}{}
\item[$^{\mathrm{a}}$] Broad component in brackets.
\end{list}
\end{table}

\subsection{Molecular hydrogen and iron}
\label{sec:h2}
The H$_2$ emission lines originate in surfaces of molecular clouds exposed to stellar or nuclear radiation. The observed line strengths and ratios strongly depend on the excitation mechanisms of H$_2$ discussed below.

[\ion{Fe}{II}] emission is believed to originate in partially ionized zones, which occur around supernovae or the active nucleus. Such regions can be produced by the hard ionizing X-ray/UV continuum or by shocks in the circumnuclear gas or in supernova remnants \citep[e.g.][]{1997ApJ...482..747A}.

\subsubsection{Kinematics of the H$_2$ and [\ion{Fe}{II}] lines}
\label{sec:kinematics}
The $H+K$ spectra exhibit a couple of rotational/vibrational lines of molecular hydrogen (Fig. \ref{fig:hk_spec}). The most prominent is the 1-0S(1) transition. Figure \ref{fig:hk_nuclear}(b) shows the spatial distribution of H$_2$, which appears rather confined to the nucleus. Most of the 1-0S(1) emission is concentrated within 260~pc (30\% peak contour in Fig. \ref{fig:hk_nuclear}(a)) around the nucleus. The emission further shows a smooth decline towards the outer regions as it follows the continuum contours. This is in agreement with previous investigations of the molecular content of samples of Seyfert galaxies, which showed that the gas is concentrated in a disk around the centers of the galaxies \citep{2003MNRAS.343..192R,2005MNRAS.364.1041R}. The [\ion{Fe}{II}] emission is oriented almost perpendicular to the H$_2$ emission and follows the recombination line gas (Figs. \ref{fig:hk_nuclear} (b) and \ref{fig:hk_extended} (a)-(c)). 
Comparison of the line width of H$_2$ with the one of the narrow lines can thus  yield constraints on the location of the molecular and narrow-line gas. We consider line widths to be spectroscopically-resolved, if the FWHM is larger than 280~km~s$^{-1}$, i.e. $\sqrt{2}$ times the spectral resolution of 200~km~s$^{-1}$. From Table \ref{tab:linefluxes} we find that the molecular hydrogen emission is slightly resolved spectroscopically and that the line widths are similar in all regions. The H$_2$ lines are also always broader than the narrow recombination lines. In contrast to this, [\ion{Fe}{II}]~1.257~$\mu$m appears to be resolved spectroscopically only on the nucleus and in regions 2 and 4. In these three regions [\ion{Fe}{II}] is also broader than the narrow recombination lines and even broader than molecular hydrogen. Apart from that its FHWM is similar to that of the narrow recombination lines. Off-nuclear [\ion{Fe}{II}] follows the patchy Pa$\alpha$ emission/Br$\gamma$/\ion{He}{I} distribution, thereby tracing star formation. On the other hand, the unusual broadening of the iron emission in regions 1, 2, and 4 might also be related to a jet-like extension (southeast-northwest direction) in the 6~cm VLA continuum map of \cite{1984ApJ...278..544U}. Figure \ref{fig:hk_nuclear} (d) displays our Pa$\alpha$ map overlayed with the radio map in contours. It is rather speculative but conceivable that a radio jet might alter the iron gas kinematics \citep[cf. discussion in][]{2003MNRAS.343..192R}.

These observations seem to confirm previous results by \cite{2003MNRAS.343..192R} and \cite{2005MNRAS.364.1041R} that the H$_2$ kinematics is decoupled from the [\ion{Fe}{II}] and narrow recombination line kinematics.
 
However, because of the combination of low inclination and limited spectral resolution, we cannot detect any considerable dynamics from line shifts.

\subsubsection{H$_2$ excitation mechanism by line ratios.}
Three distinct processes are commonly discussed: (1) UV fluorescence (non-thermal), (2) X-ray heating (thermal), and (3) shocks (thermal). The processes result in different H$_2$ responses and can thus helping distinguishing the dominant excitation process \citep[e.g.][]{1994ApJ...427..777M}.

(1) \emph{Excitation via UV-fluorescence} 
can occur via UV pumping and/or collision with H-atoms, which are controlled by temperature, density, and strength of the UV radiation field. Within a warm, high-density gas in a strong UV radiation field, thermal line ratios are found for the lower vibrational levels. In this case the lower levels are dominated by collisional excitation, whereas the higher levels are populated via UV pumping. To distinguish this scenario from pure shock excitation, observations of the high transitions are essential. In the UV excited gas, strong 2-1H$_2$ lines are expected, as well as strong lines in the $H$ band, e.g. 6-4Q(1)~1.6~$\mu$m ~or 1-0S(7)~1.75~$\mu$m \citep[cf. Table 2 of][]{1987ApJ...322..412B}. The $H$-band lines only become weak relative to the $K$-band lines in the case of very high densities and high UV intensities. Since we detected no H$_2$ lines in the $H$ band and the 2-1H$_2$ lines are only measured as upper limits, UV radiation as the main excitation mechanism seems unlikely.

(2) \emph{Excitation by X-rays or cosmic rays}. \cite{1997ApJ...481..282T}, \cite{1990ApJ...363..464D}, as well as \cite{1996ApJ...466..561M}, analyzed the excitation of H$_2$ by X-rays and cosmic rays. Obvious non-thermal line spectra from these X-ray-dominated regions (XDRs) are only observable, if the temperature is well below 1000~K and the ionization-rate per H-atom is below $10^{-15}$cm$^3$s$^{-1}$. At higher temperatures, the lower vibrational levels will be populated mainly by collisions, whereas higher ionization-rates will destroy the H$_2$ molecules. \cite{1983ApJ...269..560L} derived the relation $L_{IR}\approx 10^{-3}L_X$ between the 1-10~keV X-ray luminosity and the 1-0S(1) line luminosity for isobaric models containing a compact X-ray source. The 2-10~keV X-ray flux of Mrk~609 \citep{2002MNRAS.336..714P} is about $2.7\times 10^{-15}$~W~m$^{-2}$ and translates to a 1-0S(1) line flux of about $\sim 10^{-18}$~W~m$^{-2}$. This is comparable in magnitude to the measured 1-0S(1) flux, if summing over all regions. The analysis of the line ratios, however, does not strongly support excitation via X-rays.

(3) \emph{Thermal excitation via shock fronts}. Thermal excitation occurs through collisions with H or H$_2$. The population of the electronic ground levels in the ro-vibrational transitions represents a Boltzmann distribution. Then the temperature of the gas represents a kinetic temperature that could be as high as 2000~K or more. See for example \cite{1993ARA&A..31..373D} for a review. 

In order to test whether the [\ion{Fe}{II}] and H$_2$ emission can be caused by X-ray heating, we used the models of \cite{1996ApJ...466..561M} to estimate the emergent 1-0S(1) and [\ion{Fe}{II}] fluxes of a gas cloud illuminated by a source of hard X-rays with an intrinsic luminosity $L_X$. The cloud is at a distance $d$ from the X-ray source and has an electron density $n_e$. The cooling is then given by the effective ionization parameter $\xi_\mathrm{eff}$:
\begin{equation}
\xi_\mathrm{eff}=1.26\times 10^{-4}\frac{f_X}{n_5N_{22}^{0.9}}
\end{equation}
where $f_X$ is the incident hard X-ray flux at the distance $d$~[pc] from the X-ray source, $n_5$~[$10^{-5}$cm$^{-3}$] is the total hydrogen gas density, and $N_{22}$~[$10^{22}$cm$^2$] the AGN intrinsic attenuating column density. The value of $f_X$ can be calculated via $L_X/4\pi d^2$. \cite{1996ApJ...466..561M} calculate the emergent intensities for two gas densities, $10^5$~cm$^{-3}$ and $10^3$~cm$^{-3}$, which can be directly read off their Fig. 6(a, b). Then, the flux can be calculated using the solid angle of the cloud region, i.e. the aperture of a 10-pixel diameter (0\farcs 5 corresponding to $4.6\times 10^{-12}$~sr). The \emph{BeppoSAX} 0.1-10~keV X-ray luminosity is $6.3\times 10^{42}$~erg~s$^{-1}$ (in the high state) and the absorbing column density is $N_H\leq 1.32\times 10^{21}$~cm$^{-2}$ for a single power-law model \citep{2002MNRAS.336..714P}. For the indicidual regions, Table \ref{tab:h2_feii} lists the H$_2$ and [FeII] fluxes in units of $10^{-20}$~W~m$^{-2}$ for the distances $d$ in pc.

\begin{table}[h!]
\caption{Emergent H$_2$ and [\ion{Fe}{II}] fluxes using the models of \cite{1996ApJ...466..561M}.}
\begin{tabular}{c c c c c}
\hline
\hline
       & \multicolumn{2}{c}{$n=10^5$~cm$^{-3}$}& \multicolumn{2}{c}{$n=10^3$~cm$^{-3}$}\\
$d$  & 1-0S(1) & [\ion{Fe}{II}] & 1-0S(1) & [\ion{Fe}{II}]\\
\hline
50   & 1450.0 & 0.5     & ---      & ---\\
80   & 15.0   & 0.0     & 460.0    & 145.0\\
420  & 15.0   & 0.0     & 0.05     & 0.0\\
630  & 9.0    & 0.0     & 0.08     & 0.0\\
1300 & 3.0    & 0.0     & 0.15     & 0.0\\
\hline
\end{tabular}
\label{tab:h2_feii}
\end{table}

Taking the uncertainties in our flux calibration into account, the H$_2$ estimates show that X-ray heating can account for part of the observed 1-0S(1) flux (cf. Table \ref{tab:linefluxes}) in the nuclear region. The model also predicts the 2-1S(1) and the Br$\gamma$ fluxes. The expected line ratios approximately resemble those of the observed 2-1S(1)/1-0S(1) and 1-0S(1)/Br$\gamma$ ratios. On the other hand, [\ion{Fe}{II}]~1.644~$\mu$m~appears not to be excited by X-ray heating, except for the nuclear region in case of the lower electron density environment. This indicates that other processes like shock heating or fluorescence have to be accounted for. In fact, the NIR diagnostic diagram (Fig. \ref{fig:nirdiag}) points towards the importance of shock heating.

\begin{figure}[h!]
\begin{center}
\resizebox{\hsize}{!}{\includegraphics{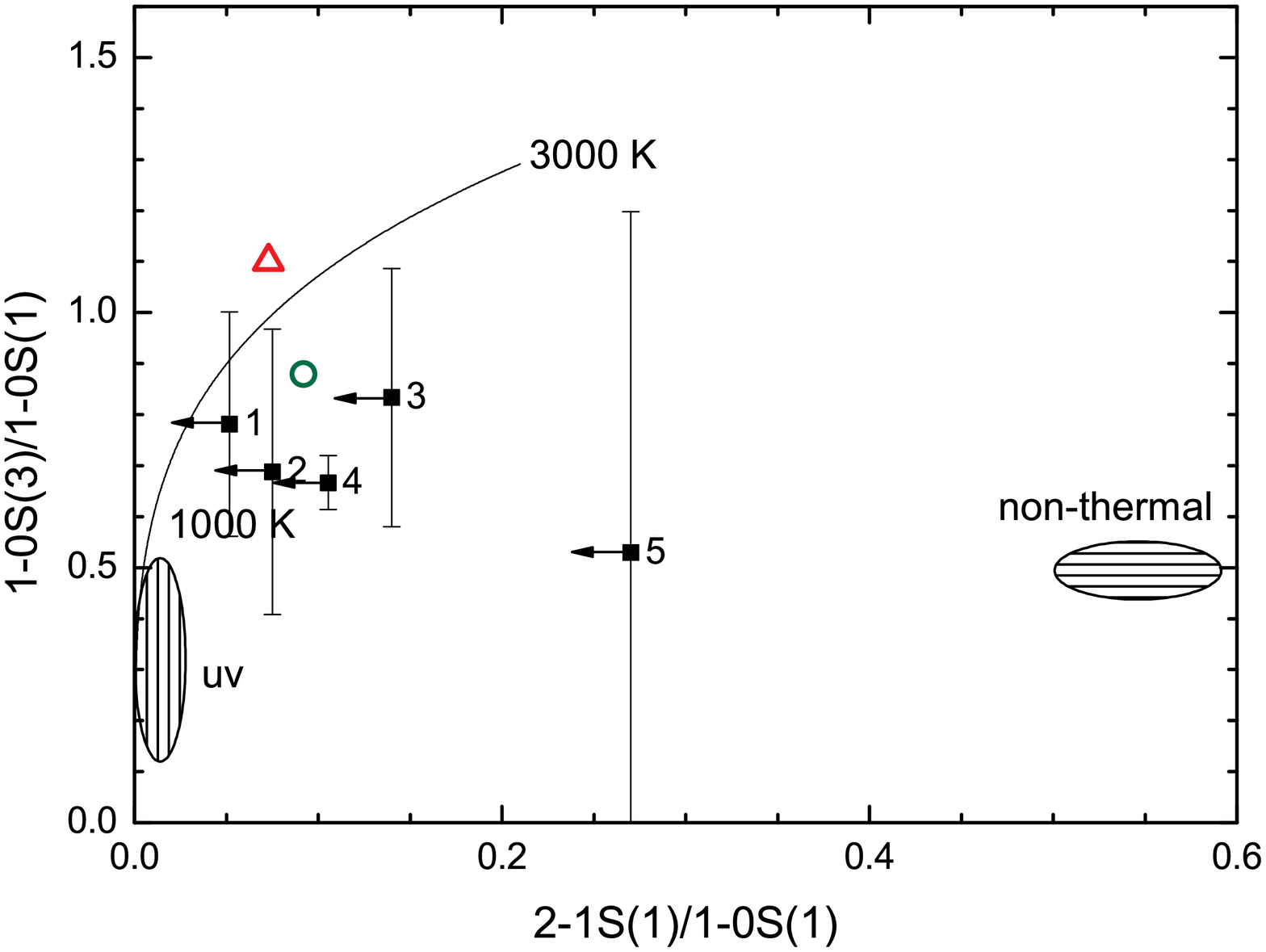}}
\caption{H$_2$ diagnostics diagram using the two line ratios 2-1S(1)/1-0S(1) vs. 1-0S(3)/1-0S(1). The region with the horizontal fill pattern covers the non-thermal models of \cite{1987ApJ...322..412B}. The vertically patterned region are the thermally uv-excited models of \cite{1989ApJ...338..197S}. The open triangle corresponds to the X-ray heating models of \cite{1990ApJ...363..464D} and the open circle to the shock model of \cite{1989MNRAS.236..929B}. Filled squares are the measured line ratios with numbers indicating the extraction region (see Fig. \ref{fig:hk_extended} (a)).}
\label{fig:h2_diagnostic}
\end{center}
\end{figure}

In summary, the observed line ratios displayed in Fig. \ref{fig:h2_diagnostic} agree quite well with those of a shock model with 2000~K from \cite{1989MNRAS.236..929B} for all regions. A scenario with mixed excitation is possible and more realistic as well. Fitting the 2-1 S(1)/1-0 S(1) ratio with the models 14 (UV excitation) and S2 (shock excitation) of \cite{1987ApJ...322..412B} for regions 1-4 about 96\% of the line emission should be due to thermal excitation and about 4\% due to UV pumping. For region 5 thermal excitation could account for up to 60\% and UV pumping for up to 40\%.

\subsubsection{The H$_2$ population diagram.}
Another approach to finding the relevant excitation mechanism is to assume thermal excitation. The ro-vibrational levels are then populated according to the Boltzmann equation. This can be illustrated by a population diagram showing the observed population density versus the energy of the upper level. If all values lie on a straight line, the gradient is proportional to the temperature of the molecular gas. The population density can be derived from the observed column density according to \cite{1988ApJ...329..641L}:
\begin{equation}
N_\mathrm{col}=\frac{f}{A_{ul}}\frac{\lambda}{hc}\times\frac{4\pi}{\Omega_\mathrm{aperture}},
\end{equation}
where $[f]=\mathrm{W~m}^{-2}$ is the observed line flux, $A_{ul}$ the transition probability \citep[taken from][]{1998ApJS..115..293W}, $\lambda$ the rest frame line wavelength, $h$ the Planck constant, and $c$ the velocity of light. The additional factor $4\pi/\Omega_\mathrm{aperture}$ takes the spatial distribution into account. The Boltzmann distribution for thermal equilibrium can be written as
\begin{equation}
\frac{N'}{N''}=\frac{g_J'}{g_J''}e^{-\Delta E/kT},
\end{equation}
where $N'$ and $N''$ denote the column densities at the corresponding levels, $g_J'$ and $g_J''$ are the statistical weights, and $T$ the temperature of the thermal equilibrium. If the H$_2$ is non-thermally excited, only lines within one rotational level (e.g. $v=1-0$) fall on a straight line.

\begin{figure}[h!]
\begin{center}
\resizebox{\hsize}{!}{\includegraphics{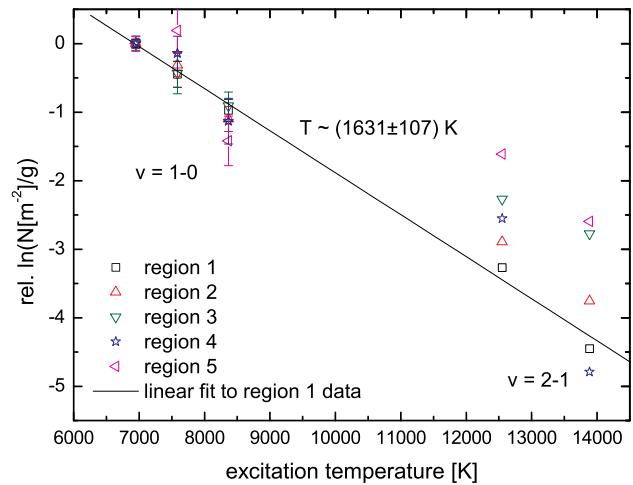}}
\caption{H$_2$ level population diagram relative to 1-0S(1) for the the five regions of Fig. \ref{fig:hk_extended} (a). All $v=2-1$ fluxes are upper limits. A linear fit to the region 1 data is presented and yields an excitation temperature of about $T_\mathrm{ex}=(1631\pm 107)$~K (cf. Fig. \ref{fig:h2_diagnostic}).}
\label{fig:h2_temp}
\end{center}
\end{figure}

The population diagram Fig. \ref{fig:h2_temp} for the H$_2$ lines of Mrk~609 shows that, within their errors, the different transitions fall on a straight line. From this we estimate a kinetic temperature for the nuclear region of about $1631\pm 107$~K. It has to be noted that the $v=2-1$ fluxes are upper limits. For region 1 these values are consistent with the $v=1-0$ kinetic temperature. For regions 2, 3, and 4, the scatter with respect to a linear relationship is larger and the $v=2-1$ values tend to show lower excitation temperatures. Nevertheless, the temperatures found in the population diagram agree with the temperatures indicated by the solid curve in Fig. \ref{fig:h2_diagnostic}. A stronger non-thermal contribution might be possible for region 5.

\begin{figure}[h!]
\begin{center}
\resizebox{\hsize}{!}{\includegraphics{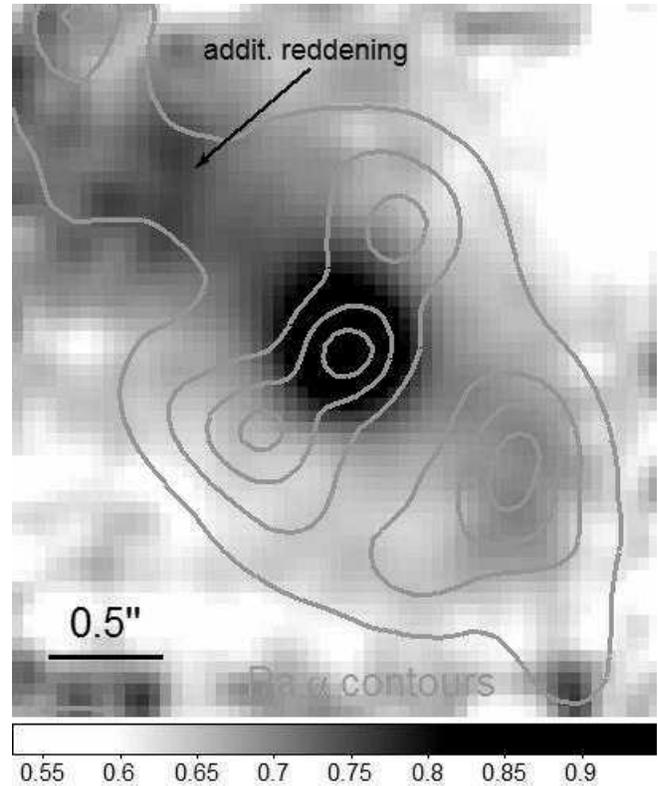}}
\caption{Color map derived from continuum fluxes measured at the 2MASS effective $H$ and $Ks$ band wavelengths and zero points. Darker greyscales indicate redder colors with $H-K$ ranging from 0.5 to 1. Contours correspond to Pa$\alpha$ emission displayed in Fig. \ref{fig:hk_extended} (a).}
\label{fig:reddening}
\end{center}
\end{figure}

\subsection{Stellar absorption features}
\label{sect:stellar}
\begin{figure}[h!]
\begin{center}
\resizebox{\hsize}{!}{\includegraphics[angle=90]{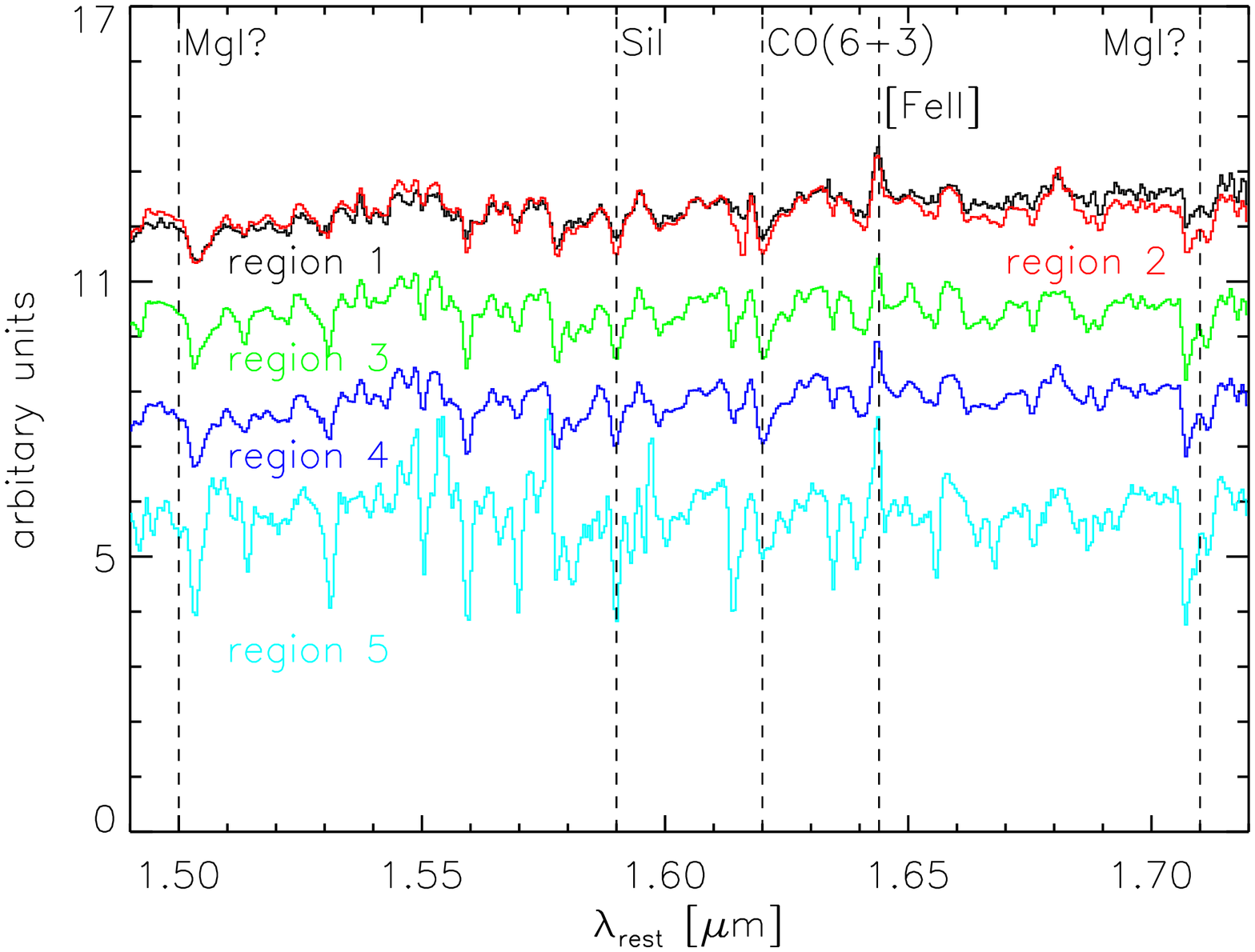}}
\resizebox{\hsize}{!}{\includegraphics[angle=90]{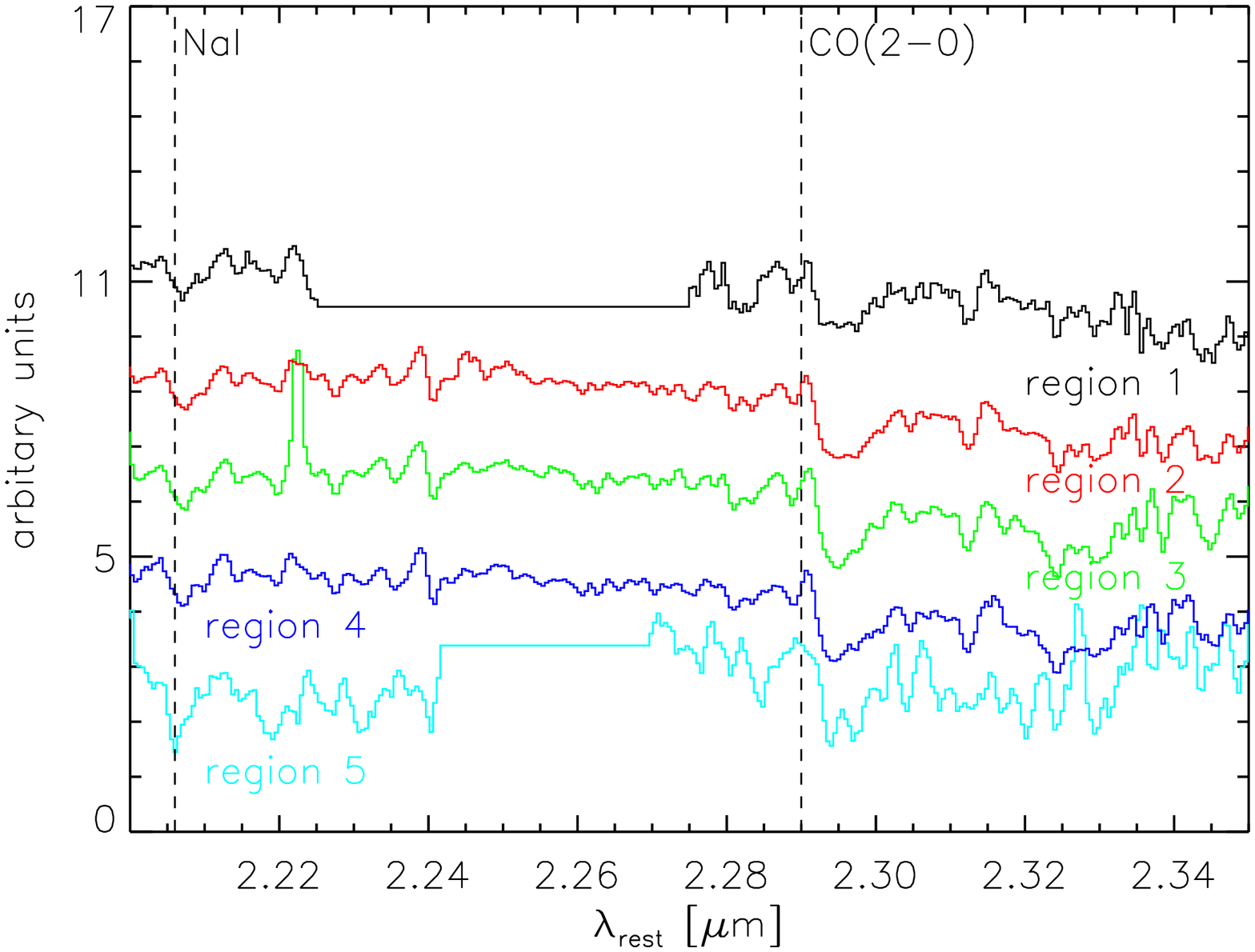}}
\caption{Stellar absorption features in the $H$-band (\emph{upper panel}) and $K$-band (\emph{lower panel}) parts of the spectra extracted from the five regions of Fig. \ref{fig:hk_extended} (a). The fluxes of the $H$ and $K$ spectra are scaled to the flux at 2.1~$\mu$m of region 1. Then the spectra were shifted by an arbitrary amount for better visibility. The $K$-band part of region 1 between 2.225 and 2.275~$\mu$m is not useable because of a detector defect and has been replaced by a flat line. The same situation applies for region 5 between 2.241 and 2.27~$\mu$m. 
}
\label{fig:hk_stellar}
\end{center}
\end{figure}

The $H+K$ spectra of all five regions show stellar absorption features like the CO(6-3)~1.62~$\mu$m~ and CO(2-0)~2.29~$\mu$m, as well as \ion{Si}{I}~1.59~$\mu$m, \ion{Mg}{I} 1.5,1.71~$\mu$m, and \ion{Na}{I}~2.206~$\mu$m (see Fig. \ref{fig:hk_stellar}). The details of the stellar content will be discussed in a subsequent paper, but one can get initial insight into the stellar populations by using the ratio of equivalent widths (EW) of CO(6-3) and \ion{Si}{I} 1.59~$\mu$m to indicate the temperature of late-type stars \citep{1993A&A...280..536O,2004ApJS..151..387I}. CO(6-3) grows rapidly from early K to late M stars, while SiI varies only very slowly with temperature. Table \ref{tab:ew} lists the ratio for all regions. 
The measured ratios correspond to M-type giants \citep[see Fig. 5(b) in][]{1993A&A...280..536O}. One has to take into account that in active galaxies these stellar features can be substantially diluted by non-stellar nuclear emission. This effect can be recognized by comparing spectra of regions 1 and 2 of both panels in Fig. \ref{fig:hk_stellar}. While being very similar in the $H$ band, the nuclear $K$-band spectrum is significantly reddened. However, we do not go on to consider the effect at this point.

\begin{table}[h!]
\caption{Measurements of stellar absorption lines.}
\begin{tabular}{cccc}
\hline
\hline
Region & EW(CO(6-3))/EW(SiI) & CO(2-0) index & EW(\ion{Na}{I})\\
       &                     &               & [\AA]\\
\hline
1      & 0.31 & 0.09 & 2.5\\
2      & 0.28 & 0.19 & 2.8\\
3      & 0.26 & 0.14 & 3.0\\
4      & 0.31 & 0.13 & 3.0\\
5      & 0.24  & 0.17 & 3.8\\
\hline
\end{tabular}
\label{tab:ew}
\end{table}

The CO(2-0) index can be used to estimate the age of the stellar population \citep{2000A&A...357...61O}. As outlined in \cite{2006ApJ...645..148R}, we use the Ivanov CO index \citep{2004ApJS..151..387I} and convert it to a photometric CO index to read the approximate age from Fig. 1 of \cite{2000A&A...357...61O}. The photometric CO indices range from 0.1 to 0.15. For solar metallicity this corresponds to ages between 80 and 150~Myr for the Geneva tracks with the younger ages measured in regions 2 and 5. Such deep CO absorption is typically found in starburst galaxies \citep[e.g.][~and references therein]{2006A&A...452..827F}. Note that the blue edge of the CO(2-0) bandhead is influenced by atmospheric OH emission, and the change in index is on the order of 20\%. Thus, the stellar ages might become shorter by a few 10~Myr.

The \ion{Na}{I} absorption is commonly found in late-type stars. Model calculations by \cite{2005ApJ...633..105D} show that the EW of \ion{Na}{I} typically ranges between 2 and 3~\AA. Their models include AGB phases that have a significant influence on the depth of the \ion{Na}{I} feature \citep[see Fig. 7 of][]{2005ApJ...633..105D}. Indeed, the EWs range between 2.5 (for region 1) and 3.8 (for region 5), which is consistent with the ages found from the CO(2-0) index. Both results agree with the young stars found by \cite{1988ApJ...332..172R} from their UV/optical spectroscopy.

\subsection{CO(1-0) observations}
Another ingredient in the starburst/Seyfert picture is the presence of fuel in the form of molecular gas. Giant molecular clouds are the birthplaces of stars. The dense gas is dissipative and galaxy interaction is believed to channel huge amounts of molecular gas towards the nucleus of the interacting galaxies, providing fuel for the nuclear activity \citep[e.g.][]{2005MNRAS.361..776S}. Infrared luminous AGN ($L_{IR}>10^{10}$~erg~s$^{-1}$) are found to be rich in molecular gas \citep[e.g.][~and references therein]{2001AJ....121.3285E,2005ApJS..159..197E} with gas masses of up to $M_{\mathrm{H}_2} \sim 10^{10} M_\odot$.
\citet{1997ApJ...485..552M} explain the more intense star formation found in Seyfert 2 galaxies with a significantly higher fraction of asymmetric morphologies than in Seyfert 1 and field galaxies. As discussed in Sect. \ref{sec:fueling}, there is compelling observational and theoretical evidence that bars efficiently redistribute angular momentum in galaxies and cause gas to flow inwards into the circumnuclear region (inner 1-2~kpc), giving rise to starburst and nuclear activity \citep{2004ARA&A..42..603K,2005ApJ...630..837J}. A major result from the Nuclei of Galaxies (NUGA) project carried out at the Plateau de Bure interferometer is the wide variety of circumnuclear disk morphologies on the 100~pc scale associated with the AGN (\citeauthor{2005A&A...441.1011G} \citeyear{2005A&A...441.1011G} and also \citeauthor{2000ApJ...533..850S} \citeyear{2000ApJ...533..850S} for an integrated CO/NIR study of NGC~1068). It is believed that on these scales, secondary perturbations modes appear to take over and are responsible for channeling gas towards the center of the galaxy. Still the scales are large compared with nuclear scales. The actual fueling of the central engine is believed to be supported by viscous flow of the material delivered by the secondary perturbations \citep{2004sgyu.conf..205D}.

\begin{figure}[h!]
\begin{center}
\resizebox{\hsize}{!}{\includegraphics{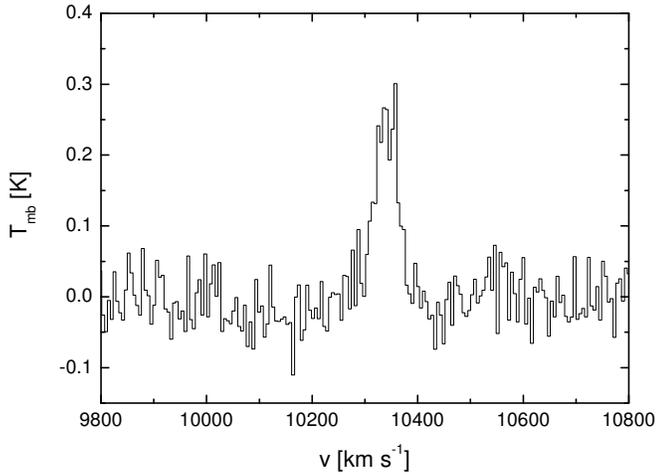}}
\end{center}
\caption{Observed-frame CO(1-0) spectrum of Mrk~609.}
\label{fig:nobeyama}
\end{figure}
During an observing campaign of cluster galaxies with the Nobeyama 45m telescope in March 2005, we also observed Mrk~609 in $^{12}$CO(1-0). The baseline-subtracted, main-beam-efficiency corrected CO(1-0) spectrum is presented in Fig. \ref{fig:nobeyama}. The beam with a size of about 15\arcsec essentially covers the entire visual part of the galaxy (cf. Fig. \ref{fig:hst}). The FWHM of the CO line is $51\pm 4$~km~s$^{-1}$ and the line-integrated flux, corrected for main-beam efficiency, is $I_\mathrm{CO}=(14\pm 1)$~K~km~s$^{-1}$. The narrow line width is reminiscent of the small inclination of Mrk~609.  According to Eq. 2 in \citet{1992ApJ...398L..29S}, the line luminosity results in $L'_\mathrm{CO}=(1.5\pm 0.1)\times 10^9$~K~km~s$^{-1}$~pc$^2$. We can now estimate the H$_2$ mass, assuming optically thick and thermalized emission, originating from gravitationally bound molecular clouds \citep{2005ApJS..159..197E}: 

\begin{equation}
\alpha=\frac{M(\mathrm{H}_2)}{L_\mathrm{CO}'}\ M_\odot\ (\mathrm{K~km~s}^{-1}\mathrm{pc}^2)^{-1}.
\end{equation}
The conversion factor $\alpha$ lies between 2 and 5 \citep{1991ApJ...368L..15R}.
Here we adopt $\alpha=4$, resulting in $M(\mathrm{H}_2)=(6.0\pm 0.4)\times10^9 M_\odot$, which is on the high end of the range of masses found in AGN \citep[e.g.][]{1997A&A...327..493R}.

The gas mass can be compared with the dust mass derived from the IRAS 60~$\mu$m and 100~$\mu$m fluxes \citep[$f_{60}=2550$~mJy, $f_{100}=4760$~mJy;][]{1985ApJ...298..614R}. The infrared emission originates from warm dust, heated either by star-formation or by the active nucleus. Following \cite{2005ApJS..159..197E}, the dust temperature results in $T_\mathrm{dust}\approx 47$~K, and the corresponding dust mass in $M_\mathrm{dust}\approx 1.1\times 10^{7}M_\odot$. The gas-to-dust ratio, $\sim 545$, is comparable to \emph{IRAS}-detected spiral and luminous infrared galaxies \citep{1991ARA&A..29..581Y}. Next, we calculate the infrared luminosity using the formula given in \citet{1997ApJ...478..144S}. \citet{2000PASJ...52..803T} furthermore show that an IRAS $K$-correction for this type of galaxy is not necessary. The color correction, $r$, lies between 1.5 and 2.1 \citep{1997ApJ...478..144S}. Assuming $r=1.8$, the infrared luminosity calculates as $L_{FIR}\approx 10^{11}L_\odot$, which is typical for an infrared luminous galaxy.  

The FIR luminosity of (inactive) galaxies is interpreted as a measure of the number of visible and UV photons, thus measuring the number of high-mass stars. This allows us to estimate the global star-formation rate (SFR) from the IRAS fluxes using Eqs. 7 and 8 in \cite{2003ApJ...599..971H}. The global SFR of Mrk~609 amounts to $\sim 30M_\odot$~yr$^{-1}$, on the other hand, CO(1-0)  measures the cold molecular gas supplying the star formation. The ratio between FIR and CO luminosity, therefore, is a measure of the star formation efficiency \citep[SFE; cf.][]{1994ApJ...424..627E}. The global SFE is on the order $L_\mathrm{FIR}/M_{\mathrm{H}_2}\approx 17\ L_\odot\ M_\odot^{-1}$, which is again on the high end of the SFEs of non-interacting galaxies. Such SFRs and SFEs are also often found in interacting systems \citep{1988ApJ...334..613S}. Stars thus appear to be formed very efficiently in Mrk~609 and the gas depletion time is short ($\sim 3\times 10^8$~yr).

We can also estimate an SFR using the 21~cm radio flux. For Mrk~609, the NRAO VLA Sky Survey (NVSS) 21~cm flux $f_{21\mathrm{cm}}=30$~mJy yields a luminosity of $L_{21\mathrm{cm}}=8.3\times10^{22}$~W~Hz$^{-1}$. According to Eqs. 1 and 2 of \cite{2003ApJ...599..971H}, the radio derived SFR amounts to $\sim 46M_\odot$~yr$^{-1}$. This is somewhat higher than the FIR-derived SFR.  \cite{2003ApJ...599..971H} find a tight correlation between radio and FIR flux. The differences in the derived SFRs could be caused by a stronger AGN contribution (e.g. jet) at the 21~cm wavelengths.
\subsection{The starburst/AGN connection in Mrk~609}
\label{sec:connection}
In contrast to previous observations of Mrk~609, we have now spatially resolved the inner circumnuclear environment on the 270~pc scale. This allows us to study the relative contributions and importance of nuclear and off-nuclear (\ion{H}{II} regions, late-type stars) emission.

\begin{figure}[h!]
\begin{center}
\resizebox{\hsize}{!}{\includegraphics{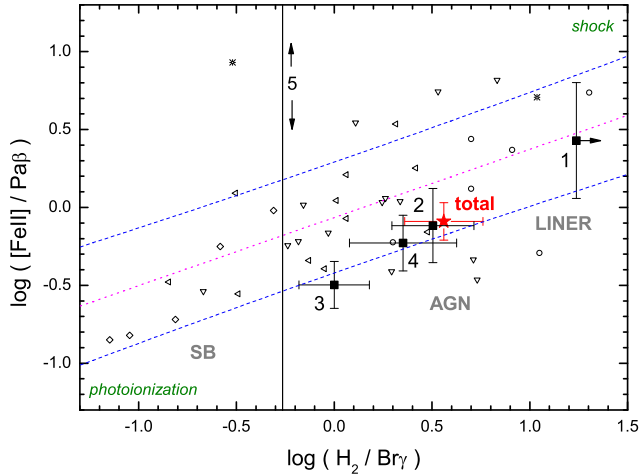}}
\end{center}
\caption{Line ratios of [\ion{Fe}{II}]~1.257$\mu$m/Pa$\beta$ and 1-0S(1)~2.121$\mu$m/Br$\gamma$. Activity types (starburst, AGN, LINER) are indicated. Filled symbols represent our measurements of the five regions of Mrk~609, as well as for the total FOV (red star). Because of missing $J$-band data, region 5 is only accurately located along H$_2$/Br$\gamma$. Open symbols correspond to literature values: rhombs represent starburst galaxies from \cite{2004ApJ...601..813D}, left triangles Seyfert 1 galaxies from \cite{2004A&A...425..457R}, headlong triangles Seyfert 2s from  \cite{2005MNRAS.364.1041R}, circles LINERs from \cite{1998ApJS..114...59L}, and asterixs supernovae also from \cite{1998ApJS..114...59L}. The diagram is interpreted as displaying the transition from pure photoionization (lower left corner) to pure shock driven (upper right) emission \citep{2005MNRAS.364.1041R}.  Assuming a linear relationship between the two line ratios, the dotted magenta line represents a linear fit to the literature data points. The blue dashed lines represent the 1$\sigma$ prediction band of the fit.}
\label{fig:nirdiag}
\end{figure}

Analogously to the optical \citep[e.g.][]{1981PASP...93....5B,1987ApJS...63..295V},  \cite{2004A&A...425..457R,2005MNRAS.364.1041R} emphasize the use of the line ratios [\ion{Fe}{II}]/Pa$\beta$ and 1-0S(1)/Br$\gamma$ to distinguish between starbursts, AGN, and LINERs (low ionization nuclear emission-line region galaxy). They find that objects with either one of the ratios being lower than 2 were predominantly identified as Seyferts. Starburst galaxies, on the other hand, are located in a region where both ratios are lower then 0.4, while LINERs have both ratios higher than 2 (Fig. \ref{fig:nirdiag}).  

In this context, the ratios describe the transition from pure shock excitation driven by supernova remnants (upper right corner of the diagram) to purely ionizing radiation powered by star formation (lower left corner). For illustration purposes we plot the data from \cite{2005MNRAS.364.1041R} as open symbols in Fig. \ref{fig:nirdiag}. Starbursts from \cite{2004ApJ...601..813D} populate the lower left region of the diagram. Seyfert 1 and 2 galaxies from \cite{2005MNRAS.364.1041R} populate the middle part, whereas LINERs \citep{1998ApJS..114...59L} can be found in the upper right corner. The emission of the supernova IC~443, on the other hand, is believed to be driven by shocks \citep{1987ApJ...313..847G}. The location of AGN (powered by X-ray heating) between these two extremes has not been fully understood yet. But most probably it is the result of a power-law continuum illuminating a slab of gas \citep[see below and ][]{2005MNRAS.364.1041R}.

It can be seen from Figs. \ref{fig:hk_extended} (b) and \ref{fig:j_band} (c) that the [\ion{Fe}{II}] emission is extended towards regions 2, 3, and 4. This extent is about perpendicular to the H$_2$ one. Moreover, the radio contours are approximately congruent with the [\ion{Fe}{II}] distribution. The NIR line-ratios measured in the five regions fall into different parts of the NIR diagnostics diagram (Fig. \ref{fig:nirdiag}). The calibration uncertainties discussed above play no role at this point, since the line ratios are calculated within a spectral band. 
Our data points clearly follow a trend from a Liner-like value at the nucleus to a starburst-like value at the most distant region 5. The circumnuclear regions 2, 3, and 4 fall in the domain of potentially mixed excitation. We discuss the situation for the individual regions below.
Under the assumption that the literature values follow a linear relationship between the line ratios \citep[from photoionization- to shock-driven excitation, see][]{1998ApJS..114...59L}, we carried out a linear regression shown in Fig. \ref{fig:nirdiag} (correlation coefficient $R\sim 0.6$ and a probability of $R=0$ of about $10^{-4}$) with its 1$\sigma$ prediction band. This fit, however, is only intended to guide the eye. A firm correlation has not been established yet. Notice that our data fall somewhat below the linear relation, but still within the prediction band. Incorrect estimations of the continuum levels during the measurement of emission line fluxes might in part be responsible for this trend. On the other hand,  M-type giants are the dominating population in the NIR (Sect. \ref{sect:stellar}), and stellar absorption can therefore influence the flux measurement of emission lines \citep[e.g.][who found that stellar absorption plays an important role in their observation]{1998MNRAS.297..624H}. The study of \cite{1998ApJS..114...59L} used larger apertures compared to our work from which they extracted spectra. In this case a considerable part of the host galaxy emission can modify the emission-line measurements. On the other hand, \cite{2005MNRAS.364.1041R,2004A&A...425..457R} use nuclear apertures of sizes comparable to our work. The latter study, however, does not consider circumnuclear stellar absorption features at the position of the emission lines,  which can influence the results.

Visual inspection of the spectra of late-type stars \citep[][~in $J$ and $H,K$,  respectively]{2000ApJ...535..325W,2004ApJS..151..387I} shows that there is no significant stellar absorption at the position of the emission lines. Atmospheric residuals might account for an uncertainty in flux on the order of 20\%, as discussed in Sect. \ref{sect:stellar}.

\begin{figure}[h!]
\begin{center}
\resizebox{\hsize}{!}{\includegraphics{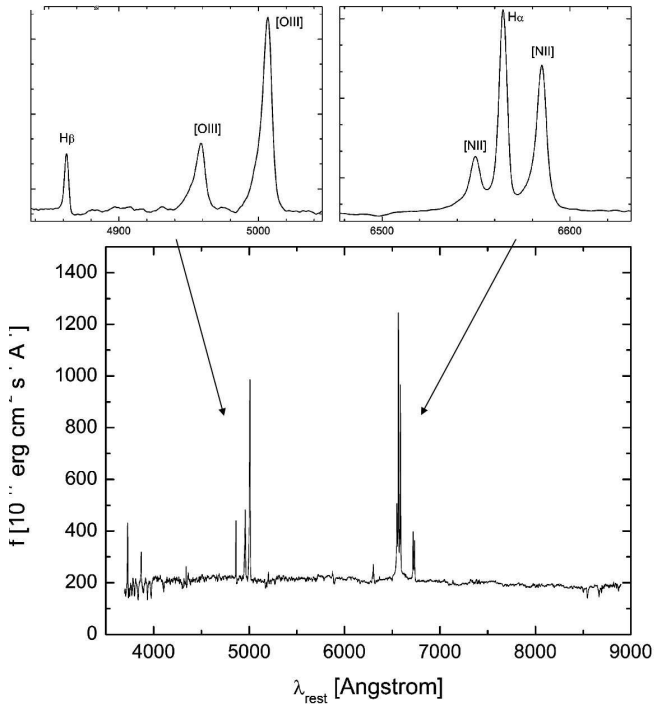}}
\end{center}
\caption{Rest frame SDSS spectrum \citep[from SDSS DR4][]{2005astro.ph..7711A}. The upper insets are close-ups of the H$\beta$ and H$\alpha$ lines. The spectra are measured through a 3\arcsec diameter fiber.}
\label{fig:sdss_spec}
\end{figure}

The nuclear region interestingly falls into the LINER regime. This contrasts with the optical classification as Sy1.8 by \cite{1981ApJ...249..462O} or from analysis of the SDSS spectrum (Fig. \ref{fig:sdss_spec}). This discrepancy may be due to aperture effects, since the SDSS spectrum has been measured in a 3\arcsec diameter fiber covering the total FOV of our NIR observations. The total narrow-line flux of the H-recombination lines is dominated by the extra-nuclear regions. The LINER character might therefore be hidden in the large-aperture optical spectrum. Such effects are also discussed by \cite{2003MNRAS.346.1055K}. This can be tested in the present case by computing the flux ratio of the line fluxes integrated over the SINFONI FOV ($\sim1.3\mathrm{kpc}\times 1.3\mathrm{kpc}$). Indeed, the position in the diagnostics diagram is shifted towards values typically found for Seyfert galaxies (Fig. \ref{fig:nirdiag}). This demonstrates the importance of high angular resolution in studying the nature of activity in galaxies. It is furthermore interesting that the off-nuclear emission regions populate a region in the diagnostic diagram where AGN are usually found. This indicates that in these regions both shock and photoionization are important.

Regions 2, 3, and 4 have line ratios typical of AGN, although the [\ion{Fe}{II}]/Pa$\beta$ ratio is close to starburst values. Region 5 has a 1-0S(1)/Br$\gamma$ ratio among those of starbursts, although the [\ion{Fe}{II}]/Pa$\beta$ ratio is unconstrained because of the lack of $J$ band  spectral information. The overlay of the Pa$\alpha$ with the 6~cm radio map (Fig. \ref{fig:hk_nuclear} (d)) from \cite{1984ApJ...278..544U} shows radio emission in all five regions. The two-point spectral index $\alpha_{6\mathrm{cm}/21\mathrm{cm}}\approx 1.3$,  assuming a power-law of the form $F_\nu\propto\nu^{-\alpha}$, is on the high end of values observed for synchrotron radiation  \citep{1984ApJ...278..544U, 2000ApJS..129...93F}. Such a steep spectral index is commonly found in star-forming late-type galaxies \citep{1992ARA&A..30..575C}. Together with the 1-0S(1)/Br$\gamma$ value, this indicates the starburst nature of the emission region 5. As mentioned above, the three other off-nuclear ratios populate the transition region from starburst to AGN in the diagram. At a projected distance of about 420~pc, regions 2 and 4 still show the strong influence of the (LINER) active nucleus. The jet-like extension of the 6~cm emission towards regions 2 and 4 might also suggest the importance of outflows and associated shocks that are able to excite [\ion{Fe}{II}] and/or H$_2$. Region 3 at a projected distance of about 630~pc lies even closer to the starburtst regime. This is expected, since the nuclear influence should decrease at larger separations.

\begin{table}[h!]
\caption{Emission line fluxes from SDSS spectrum (Fig. \ref{fig:sdss_spec}).}
\begin{tabular}{c c c}
\hline
\hline
Line & flux & FWHM\\
     & $[10^{-17}$~Wm$^{-2}$] & [km~s$^{-1}$]\\
\hline
[\ion{O}{II}]~3727\AA  & $1.9\pm 0.1$   & 480\\
H$\beta$ narrow        & $0.80\pm 0.02$ & 190\\
$[\ion{O}{III}]$~5007\AA & $7.7\pm 0.1$   & 400\\
$[\ion{O}{I}]$~6300\AA   & $0.71\pm 0.04$ & 365\\
H$\alpha$ narrow       & $4.9\pm 0.3$   & 220\\
H$\alpha$ broad        & $8.1\pm 0.4$   &3170\\
$[\ion{N}{II}]$~6585\AA  & $4.5\pm 0.3$   & 300\\
\hline
\end{tabular}
\label{tab:sdss_flux}
\end{table}

In their study of LINERs, \cite{1998ApJS..114...59L} investigated several multi-wavelength correlations. They found that LINERs and Seyferts do not follow the correlations between the IRAS 25~$\mu$m to 60~$\mu$m index and H$_2$/Br$\gamma$ and [\ion{O}{I}]/H$\alpha$ found by \cite{1992ApJ...386...68M} for starburst galaxies. The IRAS 25~$\mu$m and  60~$\mu$m fluxes are sensitive to warm (30-50~K) and hot (100-150~K) dust. Seyfert galaxies typically show a shallow index. Therefore this index can be used as a Seyfert identifier \citep{1992A&AS...96..389D}. LINERs show a wide spread in the IRAS index, whereas they only show little variation in the H$_2$/Br$\gamma$ and [\ion{O}{I}]/H$\alpha$ ratios \citep{1998ApJS..114...59L}. The values for Mrk~609 ($\alpha(25:60)=-1.9$; [\ion{O}{I}]/H$\alpha\approx 0.15$; H$_2$/Br$\gamma\approx 0.55$ for the total FOV or H$_2$/Br$\gamma\approx 1.23$ for the nuclear region) fall between the starburst correlation and the region occupied by LINERs. Note that the optical line-ratio is derived from the 3\arcsec diameter aperture SDSS spectrum, whereas the NIR line-ratio is calculated for region 1 and for the total FOV. Nevertheless, a trend towards LINER values can be recognized at both wavelengths. Besides the aperture effects mentioned above, considerable variability in the emission has been detected \citep{1988ApJ...332..172R,2002MNRAS.336..714P}, which can be combined with the \emph{duty cycle hypothesis} of \cite{1995ApJ...445L...1E}. The authors propose recurring accretion events that enhance the non-stellar continuum and generate Seyfert characteristics like broad emission lines. After the accretion event, the ionizing flux drops and the high-ionization states weaken, whereas the lower-ionization lines persist much longer, because of longer decay and light crossing times compared to the broad-line region. In this respect, Mrk~609 might approach a quiescent activity state in which the relative contribution of lower-ionization and shock-driven emission is enhanced with respect to photoionization. The starburst, since farther away from the nucleus, remains constant. Unfortunately, no multi-epoch/multi-wavelength observations of Mrk~609 exist that could substantiate this scenario; and due to the calibration problems we cannot compare our line fluxes with the NIR measurements of \cite{1990ApJ...355...88G} and \cite{1990ApJ...363..480R}.

\begin{figure}[h!]
\begin{center}
\resizebox{\hsize}{!}{\includegraphics{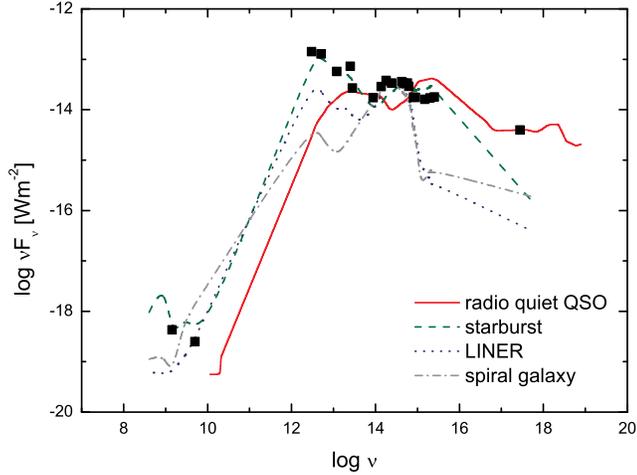}}
\end{center}
\caption{Spectral energy distribution of Mrk~609 constructed from literature values \citep{1988ApJ...332..172R,1996ApJS..106..399P}. The values are compared with average SEDs from normal galaxies, starbursts, LINERs \citep{1997AJ....114..592S}, and radio-quiet QSOs \citep{1994ApJS...95....1E}.}
\label{fig:sed}
\end{figure}

The panchromatic properties of Mrk~609 reveal its composite nature. In Fig. \ref{fig:sed} we plot the spectral energy distribution (SED) and compare it to average SEDs of starburst, LINER, and spiral galaxies \citep[taken from][]{1997AJ....114..592S}. Also shown is the average SED of radio-quiet quasars from \cite{1994ApJS...95....1E}. It can be seen that the X-ray flux is nicely represented by an AGN-like contribution. \cite{2002MNRAS.336..714P} estimate from the FIR luminosity that a starburst can only contribute up to 50\% of the X-ray luminosity. But the AGN itself cannot explain the longer wavelength fluxes, which appear to be a combination of the three other contributions. Only starbursts can produce the strong FIR fluxes. The LINER SED also fits nicely into this picture. Note that Mrk~609 is missing a significant big blue bump, which is also absent in low-luminosity AGN and might be explained by low accretion-rate models \citep{1999ApJ...516..672H}. Furthermore, \cite{1988ApJ...324..767W} found a strong correlation between the broad H$\alpha$ and the 2-10~keV X-ray luminosity as a measure for the ionizing strength of the active nucleus. \cite{2002MNRAS.336..714P} find, that the H$\alpha$ emission from the large aperture spectrum is underluminous by a factor of about 40 and suggest a deficit of ionizing photons. This is agrees with the missing UV-bump in the SED. Another possibility is the variability of the nonstellar continuum over the 15 years between optical and X-ray emission. 
 
The nuclear continuum and emission-line morphology is not symmetric and possibly resembles that of a weak nuclear bar (Fig. \ref{fig:hk_extended}(d), \ref{fig:j_band}(a)).
The emission peaks of the Pa$\alpha$ map are aligned with the major and minor axes of the bar. In addition, the slight elongation of the 6~cm radio emission towards regions 2 and 4 (Fig. \ref{fig:hk_nuclear}(d)) could be interpreted as a jet, which might in part be responsible for the heating of regions 2 and 4 \citep[e.g.][]{1993ApJ...414..563V}. The interpretation is still not clear from our data. 
On the other hand, the clumpy structure of the ISM seen in the recombination line maps might also resemble a ring-like structure or parts of a nuclear spiral. Such structures have been uncovered in both the observation \citep[e.g.][]{2002AJ....124...65E,2003ApJ...589..774M,2003MNRAS.345.1297E} and simulation \citep[e.g.][]{2000ApJ...528..677E,2004MNRAS.354..892M} of several quiescent and active galaxies. Torques associated with the gas response to the stellar-bar potential allow considerable amounts of gas to flow closer to nuclear regions \citep[$\sim 1 M_\odot$~yr$^{-1}$ on time scales of $\sim 10^8$ yr;][]{2004ApJ...617L.115E} where viscous processes take over the dissipation of angular momentum \citep{2004sgyu.conf..205D}.
From the symmetrical appearance with respect to the bar's minor axis, we might expect a northeastern counterpart for region 3, which is missing, however, in the hydrogen emission-line maps. Nevertheless, the reddening map exhibits some additional reddening at the position of the proposed counterpart. The velocity maps of the hydrogen recombination line do not show any dynamical behavior, which is expected due to the face on view of the galaxy and the limited spectral resolution. Therefore, we are currently not in a position to favor any of the above models.

\section{Summary and conclusions}
\label{sec:summary}
Mrk~609 was chosen from a sample of AO-suitable optical counterparts of luminous ROSAT X-ray AGN at low to intermediate redshifts \citep{2004ASPC..311..325Z,scienceWithAO05}. It is one of the lowest redshift objects in the sample and is therefore best-suited for initial integral field observations.

We have presented first results on the circumnuclear structure of the starburst/Seyfert composite Mrk~609 in the NIR using imaging spectroscopy with SINFONI. The morphology is complex, and the continuum images reveal a bar-like structure. The distribution of hydrogen recombination emission (Pa$\alpha$, Pa$\beta$, Br$\gamma$) is clumpy and peaks at the tip where the potential bar meets the spiral arms and in regions along the minor axis. Bars can account for the angular momentum transfer necessary for fueling nuclear activity. The 6~cm VLA emission is also extended in the latter direction. Whether this emission is associated with a jet or with resonances in the bar potential, is unclear. The presence of nuclear broad Pa$\alpha$ and [\ion{Si}{VI}] are clear indicators of the accretion of matter onto a nuclear super-massive black hole. 

The analysis of molecular hydrogen and [\ion{Fe}{II}] emission indicates the importance of shock heating, although X-ray heating by the nucleus and non-thermal contributions is possible. The distribution of molecular hydrogen follows the continuum shape, while that of [\ion{Fe}{II}] is aligned with the minor axis of the continuum and with the radio contours, as well as with the H-recombination line peaks. The nucleus itself shows signs of LINER activity, which can be recognized by its high [\ion{Fe}{II}]/Pa$\beta$ and H$_2$/Br$\gamma$ values. Our integral-field data clearly resolve the nuclear and starburst activity in the central kilo parsec. Extinction appears to play no crucial role in this region, since the H-recombination line ratios are consistent with unreddened case-B values. This is supported further by the strong Ly$\alpha$ emission. However, already small amounts of extinction have considerable effects at visible wavelengths, as can be seen in the flocculent morphology of Fig. \ref{fig:hst}. Continuum reddening in the NIR, on the other hand, can already be caused by small amounts of warm/hot dust. The reddening of the continuum found toward the nucleus and the other regions is typical of AGN and star-forming regions where the dust reprocesses UV radiation from hot stars or the active nucleus. 

We found large amounts of cold molecular gas, which provides the fuel for the star-forming activity. M-type giants are the dominating stellar population in the NIR, which indicates starburst ages of about 100~Myr (for a single burst). The nuclear stellar absorption features are considerably diluted by the non-stellar continuum.

Our LINER classification, together with the published variability of the non-stellar NIR emission, might be explained by the duty-cycle hypothesis in which short-lived accretion events occur periodically and lead to the appearance of Seyfert features in the high state and of low-ionization (shock driven) emission features in the low state. 

The X-ray emission can account for some of the H$_2$ excitation. This might also explain the higher H$_2$/Br$\gamma$ ratios for the off-nuclear regions than expected for starbursts. The [\ion{Fe}{II}] emission, however, does not seem to be excited by X-rays. Shock excitation along the southeast/northwest axis might be caused partly by a radio jet impinging into the ISM. The main driver for both species therefore appears to be shock excitation.

Our results show that high spatial resolution is vital for dissecting the nuclear and starburst activity in AGN. Morphological peculiarities, such as nuclear bars and nuclear starburst rings, are best detected at NIR wavelength, since the mass dominating stellar populations have their emission maximum in the NIR, and dust extinction is smaller than at visible wavelengths. In conjunction with high spectral resolution, it is possible to trace the stellar and gaseous nuclear kinematics in order to find important constraints for dynamical models describing the infall of matter into the nuclear regions \citep[e.g.][]{2005MNRAS.364..773F}. 
This is a prerequisite for the study of composite systems in which nuclear star formation and Seyfert activity occur at similar levels of intensity.
\begin{acknowledgements}
The authors kindly thank the anonymous referee for fruitful comments and suggestions.
This work was supported in part by the Deutsche Forschungsgemeinschaft
(DFG) via grant SFB 494.
\end{acknowledgements}

\bibliographystyle{aa}
\bibliography{mrk609}
\end{document}